\documentclass[aps,longbibliography,twocolumn,superscriptaddress,groupedaddress]{revtex4-2}
\usepackage{amssymb} \usepackage{color,graphicx} \usepackage{amsmath}
\usepackage{amsbsy} \usepackage{amsthm} \usepackage{bbm}
\usepackage{bm,bbm} \usepackage{float} \usepackage{braket}
\usepackage{placeins}
\usepackage[colorlinks=true,citecolor=blue,linkcolor=red,urlcolor=red]{hyperref}
\usepackage[sort&compress]{natbib}
\usepackage{comment}
\usepackage{mathtools}

\newcommand{\bq}{\begin{equation}} \newcommand{\eq}{\end{equation}}
\newcommand{\bqali}{\begin{equation}\begin{aligned}}
\newcommand{\eqali}{\end{aligned}\end{equation}}

\graphicspath{{../Figures/}}

\newcommand{\dd}{\text{d}}
\newcommand{\xtilde}{{\raise.17ex\hbox{$\scriptstyle\sim$}}}
\providecommand{\norm}[1]{\lVert#1\rVert}
\providecommand{\ket}[1]{\lvert #1 \rangle}
\providecommand{\ave}[1]{\langle #1 \rangle}
\providecommand{\bra}[1]{\langle #1 \lvert}
\begin{document}

\author{J. L. Gaona-Reyes}
\affiliation{Department of Physics, University of Trieste, Strada Costiera 11, 34151 Trieste, Italy}
\affiliation{Istituto
Nazionale di Fisica Nucleare, Trieste Section, Via Valerio 2, 34127 Trieste,
Italy}

\author{D. G. A. Altamura}
\affiliation{Department of Physics, University of Trieste, Strada Costiera 11, 34151 Trieste, Italy}
\affiliation{Istituto
Nazionale di Fisica Nucleare, Trieste Section, Via Valerio 2, 34127 Trieste,
Italy}

\author{A. Bassi}
\affiliation{Department of Physics, University of Trieste, Strada Costiera 11, 34151 Trieste, Italy}
\affiliation{Istituto
Nazionale di Fisica Nucleare, Trieste Section, Via Valerio 2, 34127 Trieste,
Italy}	

\title{Theoretical Limits of Protocols for Distinguishing Different Unravelings}

\date{\today}
\begin{abstract}

Stochastic unravelings of Lindblad-type master equations, such as stochastic Schr\"odinger equations (SSEs), provide powerful tools to model open quantum systems and continuous measurement processes. The same master equation can be unraveled in different ways; while these unravelings differ at the level of quantum trajectories, by construction they all yield the same averaged dynamics for the density operator. A recent question of both foundational and practical relevance is whether such unravelings can be operationally distinguished, given that certain nonlinear quantities—such as covariances and higher-order moments of conditional expectation values—are  unraveling-dependent. We show that these quantities cannot be accessed unless the measurement scheme (i.e., the unraveling) is known in advance. This renders any operational protocol to distinguish unravelings fundamentally unfeasible. We further establish that assuming access to such nonlinear quantities without prior knowledge of the unraveling would enable superluminal signaling, violating relativistic causality. 

\end{abstract}
\maketitle

\section{Introduction}

In quantum mechanics, especially in the theory of open quantum systems \cite{breuer2002theory,Rivas2012}, the information about a system of interest is encoded in the statistical operator $\hat{\rho}_t$. 
The seminal works of V. Gorini, A. Kossakowski, G. Sudarshan \cite{Gorini1976}, and G. Lindblad \cite{Lindblad1976} identified, under general assumptions, its dynamical evolution; limiting to a single Hermitian Lindblad operator $\hat{L}$, the GKLS master equation reads
\begin{equation} \label{mastereq}
\frac{\dd \hat{\rho}_t}{\dd t}=-\frac{i}{\hbar}[\hat{H},\hat{\rho}_t]
+\lambda \left(\hat{L}\hat{\rho}_t \hat{L}-\frac{1}{2}\{\hat{L}^2,\hat{\rho}_t \}\right), 
\end{equation}
where $\hat{H}$ is the Hamiltonian of the system, and $\lambda$ is a positive coupling constant. 

Over the years, interest has increased in {\it stochastic unravelings} of the master equation~\eqref{mastereq}. These are suitably modified Schr\"odinger equations with the addition of stochastic terms, whose random solutions $|\psi_{\omega,t}\rangle$, often called {\it quantum trajectories} \cite{carmichael2009open}, are such that the conditional state \cite{Cao2019} $\hat{\rho}_{\omega,t}=\ket{\psi_{\omega,t}}\bra{\psi_{\omega,t}}$,  when averaged  over the noise $\omega$, is a solution to the master equation above, i.e. ${\mathbb E}_\omega [\hat{\rho}_{\omega,t}] = \hat{\rho}_t$, where ${\mathbb E_\omega}[\cdot]$ denotes the stochastic average. 

As a notational remark, in the following, we will use the term {\it conditional} to refer to any quantity, such as the state or the quantum expectation value of an observable, {\it before} taking the stochastic average; otherwise, that quantity is meant to be averaged over the noise. Moreover, we will omit indicating explicitly the dependence of a conditional quantity on the noise ($\omega$), unless confusion may arise.

Stochastic Schr\"odinger equations (SSEs) appear in a variety of contexts \cite{wiseman2009quantum,wiseman1993quantum,carmichael2009open,dalibard1992wave,nielsen2010quantum,preskill2018quantum,breuer2002theory,gisin1992quantum,molmer1993monte,daley2014quantum,Diosi1988,Ghirardi1990,Kafri2014,Tilloy2016,Tilloy2017,weiss2012quantum,hanggi1990reaction,Ghirardi1990b,Pearle1997,Wiseman2001,Adler2007,Caiaffa2017,jacobs2014quantum,wiseman1996quantum,carmichael2007quantum,piccitto2022entanglement,Cao2019,vovk2024quantum,Piccitto2024,bassi2006geometric,Genoni2016, Wu2024, Piñol2024,albarelli2017ultimate,magrini2021real,zhang2017prediction,belenchia2020entropy,zhang2017quantum}. In quantum optics, they are used to describe the dynamics of systems under continuous measurement \cite{wiseman2009quantum,wiseman1993quantum}, for different photon detection schemes~\cite{carmichael2009open,dalibard1992wave}. 
In quantum computation, they are employed to analyze the impact of noise and errors on qubits, in light of fault-tolerant quantum computing~\cite{nielsen2010quantum,preskill2018quantum}.
Within the theory of open quantum systems, they allow us to describe the reduced system dynamics at the wave function level  \cite{breuer2002theory,gisin1992quantum}.
Numerical simulations based on SSEs, including quantum Monte Carlo methods,  enable the simulation of larger quantum systems, making SSEs a common tool for tackling computationally intensive problems \cite{molmer1993monte,daley2014quantum}.
Stochastic Schr\"odinger equations are also used in the foundations of quantum mechanics to model the spontaneous collapse of the wave function \cite{Diosi1988,Ghirardi1990,Diosi2015,Diosi2018}; more recently, they have been used to model the coupling between classical gravity and quantum matter \cite{Kafri2014,Tilloy2016,Tilloy2017} offering an alternative to quantum gravity. 

The association between master equations and stochastic unravelings, which is the subject of this work, is not unique: in general, there are infinitely many ways to unravel a master equation via SSEs \cite{Ghirardi1990b,Pearle1997,Wiseman2001,Adler2007,Caiaffa2017}, as also shown later on. One might then wonder why and how to prefer one unraveling over another. From a purely mathematical point of view, a specific unraveling might be more suitable for performing either analytical or numerical calculations \cite{Gardiner1992,Kondov2003,Li2013,Donvil2023}. From a physical point of view, when SSEs are used to model quantum measurements, different unravelings encode different measurement procedures. For example, in quantum optics, the detection of photons leaking out of an optical cavity is described by a stochastic equation, where the random outcome is encoded in discrete Poisson increments~\cite{Jacobs2010,jacobs2014quantum}.  In homodyne detection, a photodetector is used to perform a Gaussian continuous measurement of an optical mode; the resulting homodyne stochastic equation is defined in terms of a Wiener process \cite{wiseman1996quantum,carmichael2007quantum,jacobs2014quantum}. These are two different measurement processes; yet at the statistical level, when averaged over the noise, they are indistinguishable from each other \cite{jacobs2014quantum}.

Within the context of stochastic dynamics, quantities such as the mean value or the variance of observables clearly play a crucial role. In the case of SSEs, there are two types of such quantities: those such as $({\mathbb E}_\omega[\langle \hat O \rangle_t])^n \equiv (\text{Tr}[\hat O \hat{\rho}_t])^n$ (since $\hat{\rho}_t = {\mathbb E}_\omega[|\psi_t\rangle \langle\psi_t|]$), which, by construction, can be read directly from the density matrix without the need of the unraveling, except possibly for numerical reasons, as mentioned before; and those like  ${\mathbb E}_\omega[\langle \hat O \rangle_t^n]$ ($n > 1$), which depend nonlinearly on the conditional state of the system $\hat{\rho}_{\omega,t}$ so they cannot be derived from the density matrix. The latter quantities crucially depend on the chosen unraveling. The typical example of this second type of quantities is ${\mathbb E}_\omega [\Sigma_t(\hat{O}_i,\hat{O}_j)]$, with $\Sigma_t(\hat{O}_i,\hat{O}_j)=\frac{1}{2}\ave{\{\hat{O}_i, \hat{O}_j\}}_t-\ave{\hat{O}_i}_t\ave{\hat{O}_j}_t$, which for example appear in quantum optics \cite{Genoni2016, Wu2024, Piñol2024,albarelli2017ultimate,magrini2021real,zhang2017prediction,belenchia2020entropy,zhang2017quantum} in the form of the quantum Riccati equation  for the covariance matrix for the quadratures of a mechanical system in the limit of Gaussian states.

The fact that some quantities, which are nonlinear in the conditional state $\hat{\rho}_{\omega,t}$, are unraveling-dependent is known in the literature \cite{Bassi2005,Cao2019,Vovk2022}, although previous works have not focused on the possibility of uniquely identifying these unravelings at an operational level. Recently, it has been suggested~\cite{Piñol2024} that these nonlinear quantities can be used to {\it operationally distinguish} different unravelings of a given master equation. This question is both interesting and relevant.  

We argue that nonlinear quantities can be operationally computed only if the measurement procedure is known beforehand; as such, they cannot be used to operationally distinguish different measurement procedures. If the measurement scheme---or, equivalently, the unraveling---is not given, they are not accessible. The latter point is particularly crucial, as their hypothetical accessibility without knowing the measurement procedure would enable a protocol for superluminal signaling, which is (supposed to be) forbidden.

\begin{figure}[t!] 
    \centering
     \includegraphics[width=1\linewidth]{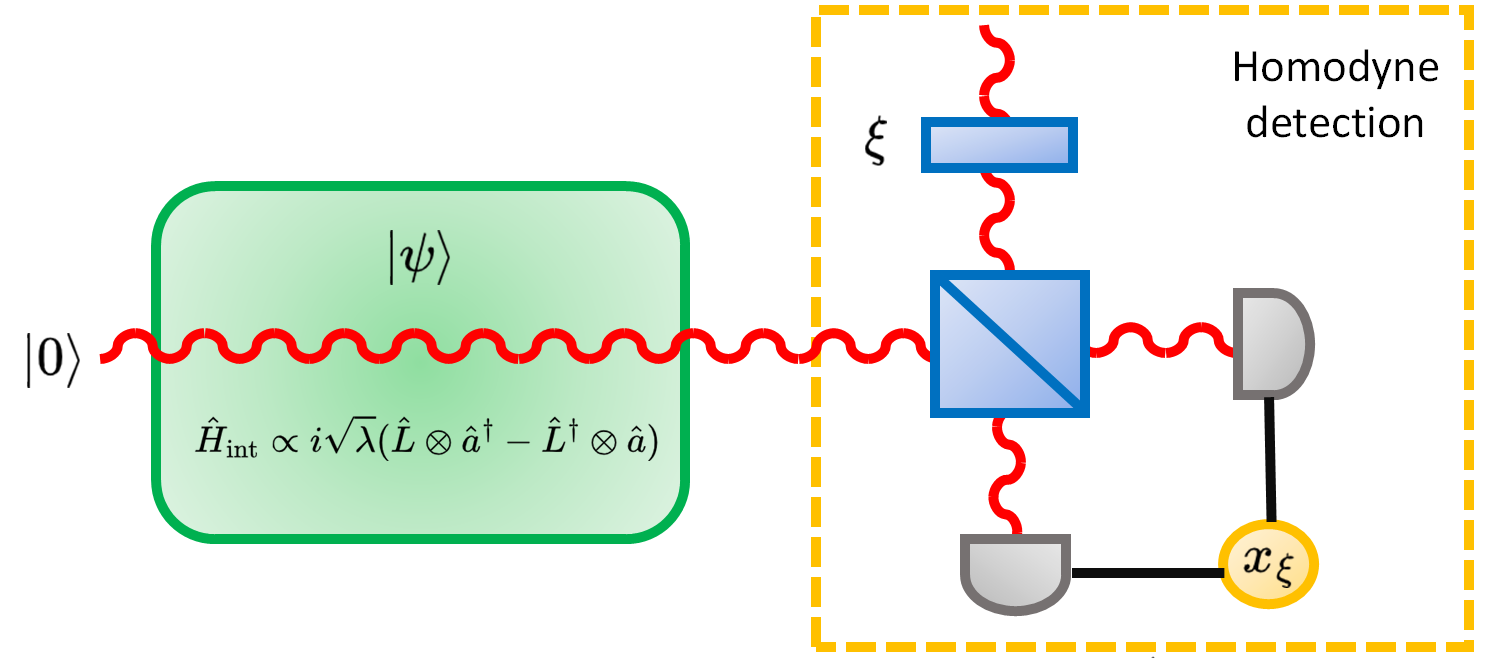}
    \caption{Indirect monitoring of the observable $\hat{L}$. The system of interest, with wave function $\ket{\psi}$, couples to the ground state $\ket{0}$ of a bath of bosonic modes with creation and annihilation operators $\hat{a}$ and $\hat{a}^\dagger$, through the interaction Hamiltonian $\hat{H}_{\text{int}}$. The light-field of the bath goes through homodyne detection, which produces the outcome $x_\xi$, where the phase $\xi$ can be controlled through a phase plate, and the evolved state is projected into the eigenstate $\ket{x_\xi}$ of the operator $\hat{x}_\xi=(\xi \hat{a}+\xi^* \hat{a}^\dagger)/\sqrt{2}$. The stochastic differential equation for $\ket{\psi_t}$ in Eq.~\eqref{unrav} arises from normalizing the state after projection into the eigenstate $\ket{x_\xi}$ of $\hat{x}_\xi$, and noticing that the outcome $x_\xi$ can be treated as a stochastic variable.}
    \label{homfig}
\end{figure} 

\section{Homodyne-detection unravelings of the master equation}

There exists a large variety of unravelings of a master equation such as Eq.~\eqref{mastereq}; for the purposes of our work, in the following we will consider the following family of continuous unravelings \cite{Adler2007}:
\begin{equation} \label{unrav}
\begin{split}
\dd \ket{\psi_t}&\!=\!\left(\!-\frac{i}{\hbar} \hat H dt \!-\!\frac{\lambda}{2}(|\xi|^2 \hat{L}^2-2 \xi \xi^\text{\tiny{R}} \hat{L}\ave{\hat{L}}_t+(\xi^\text{\tiny{R}})^2 \ave{\hat{L}}_t^2)\dd t \right.\\
&\left.+\sqrt{\lambda} (\xi \hat{L} - \xi^\text{\tiny{R}} \ave{\hat{L}}_t)\dd W_t\right)\ket{\psi_t},
\end{split}
\end{equation}
where $\ave{\hat{L}}_t=\bra{\psi_t}\hat{L} \ket{\psi_t}$, $W_t$ is a standard Wiener process and $\xi=\xi^\text{\tiny{R}}+i \xi^\text{\tiny{I}}$ is a complex parameter such that $|\xi|=1$.  Throughout this work we will use the It\^o formalism. 
As discussed in Appendix \ref{HomApp}, the above family of unravelings corresponds to an indirect continuous monitoring of the observable $\hat{L}$ through homodyne detection. Fig. \ref{homfig} shows a schematic representation of the process.

In this continuous monitoring process, the measurement record reads:
\begin{equation} \label{measrec}
\dd y_t =  \xi^{\text{\tiny{R}}} \ave{\hat{L}}_t\dd t+ \frac{1}{2 \sqrt{\lambda}}\dd W_t.   
\end{equation}
From a known initial state, using Eqs.~\eqref{measrec} and \eqref{unrav} one can  compute the state at later times: specifically, starting from the state $|\psi_t\rangle$ at a given time $t$, one  computes the conditional expectation $\ave{\hat{L}}_t$. By measuring the signal $y_t$, Eq.~\eqref{measrec} allows one to reconstruct the noise contribution $\dd W_t$, which can be used to fully determine the state at a time $t+\dd t$ via Eq.~\eqref{unrav}. 

The stochastic equation for the conditional density operator $\hat{\rho}_{\omega,t}$ reads:
\begin{equation} \label{eq:cme}
\begin{split}
\dd \hat{\rho}_{\omega,t}&=-\frac{i}{\hbar}[\hat{H},\hat{\rho}_{\omega,t}]\dd t - \frac{\lambda}{2}[\hat{L},[\hat{L},\hat{\rho}_{\omega,t}]]\dd t \\
&+\sqrt{\lambda}(\xi^{\text{\tiny{R}}} (\{\hat{L},\hat{\rho}_{\omega,t} \}-2 \ave{\hat{L}}_t \hat{\rho}_{\omega,t})\,+\,i \xi^{\text{\tiny{I}}} [\hat{L},\hat{\rho}_{\omega,t}])\dd W_t,
\end{split}
\end{equation}    
and that  for the quantum expectation value $\ave{\hat{O}}_t$ of an observable $\hat{O}$ reads: 
\begin{equation} \label{evunrav}
\begin{split}
&\dd \ave{\hat{O}}_t=-\frac{i}{\hbar}\ave{[\hat{O},\hat{H}]}_t\dd t-\frac{\lambda}{2}\ave{[\hat{L},[\hat{L},\hat{O}]]}_t\dd t\\
&+\sqrt{\lambda}(\xi^{\text{\tiny{R}}}(\ave{\{\hat{O},\hat{L} \}}_t-2 \ave{\hat{O}}_t\ave{\hat{L}}_t)+i \xi^{\text{\tiny{I}}}\ave{[\hat{O},\hat{L}]}_t)\dd W_t\\
&\equiv A_{\psi_t}\dd t + B_{\psi_t}\dd W_t.
\end{split}
\end{equation}
Equation~\eqref{eq:cme} reduces to the Lindblad equation~\eqref{mastereq} when the average over the noise is taken, since the second line averages to 0.  Similarly, Eq.~\eqref{evunrav} implies that $\mathbb{E}_\omega[\ave{\hat{O}}_t]$ does not depend on the chosen unraveling and can be computed directly from the density matrix $\hat \rho_t$.

However, as mentioned before, quantities such as $\mathbb{E}_\omega[\ave{\hat{O}}_t^n]$ with $(n>1)$, are unraveling-dependent, since 
\begin{equation} \label{unravn}
\begin{split}
\dd \ave{\hat{O}}_t^n&=\left(n \ave{\hat{O}}^{n-1}_t A_\psi + \frac{1}{2}n(n-1) \ave{\hat{O}}^{n-2} B_\psi^2 \right)\dd t\\
&+ n \ave{\hat{O}}^{n-1}_t B_\psi \dd W_t,
\end{split}
\end{equation} 
with $A_{\psi}$ and $B_{\psi}$ defined in Eq.~\eqref{evunrav}. When calculating $\mathbb{E}_\omega[\ave{\hat{O}}_t]$, a contribution proportional to $B_{\psi}^2$, which is unraveling-dependent, remains.

Thus far, we have  specified  neither the Hamiltonian nor the Lindblad operator $\hat{L}$. Thus, the unraveling dependence of quantities such as $\ave{\hat{O}}_t^n$ can be explicitly shown in any system where stochastic unravelings are used to describe its dynamical evolution. For illustrative purposes, we specialize on levitated optomechanical systems in the following section. The interested reader can find other simpler examples of the unraveling-dependence of $\ave{\hat{O}}_t^n$ in Appendix~\ref{AppExamples}.

\section{Homodyne detection of the position of a levitated nanoparticle}

As an example of the unraveling-dependence of quantities which are nonlinear in the conditional state $\hat{\rho}_{t,\omega}$, let us consider a trapped levitated nanoparticle undergoing  homodyne detection of its position. For this system, we will be interested in explicitly showing the unraveling-dependence of the components $\Sigma_t(\hat{O}_i,\hat{O}_j)$ of the covariance matrix of the system. 

A realistic dynamical evolution of the particle should include dissipation effects \cite{Setter2018,Wu2024}. Therefore, we consider the following conditional evolution for $\hat{\rho}_{\omega,t}$
\begin{equation} \label{levpart}
\begin{split}
&\dd \hat{\rho}_{\omega,t}=-\frac{i}{\hbar}\left[\frac{\hat{p}^2}{2m}+\frac{1}{2}m \Omega^2 \hat{x}^2,\hat{\rho}_{\omega,t} \right]\dd t\\
&+ \mathcal{D}_{\text{CL}}[\hat{\rho}_{\omega,t}]\dd t+ \mathcal{D}_{\text{th}}[\hat{\rho}_{\omega,t}]\dd t- \frac{\lambda}{2}K[\hat{x},[\hat{x},\hat{\rho}_{\omega,t}]]\dd t\\
&+ \frac{\sqrt{\lambda}}{\sqrt{K}} \left(\xi^{\text{\tiny{R}}}(\{ \hat{x},\hat{\rho}_{\omega,t} \}-2 \ave{\hat{x}}\hat{\rho}_{\omega,t})+i \xi^{\text{\tiny{I}}}K[\hat{x},\hat{\rho}_{\omega,t}] \right)\dd W_t.
\end{split}
\end{equation}
The Hamiltonian driving the unitary part of the evolution is that of a harmonic oscillator of frequency $\Omega$. $\mathcal{D}_{\text{CL}}[\hat{\rho}_{\omega,t}]$ and $\mathcal{D}_{\text{th}}[\hat{\rho}_{\omega,t}]$ are the modified Caldeira-Leggett dissipator and the thermalisation dissipator, respectively. The former corresponds to: 
\begin{equation}
\begin{split}
\mathcal{D}_{\text{CL}}[\hat{\rho}_{\omega,t}]=&-\frac{i \gamma}{2 \hbar}[\hat{x},\{\hat{p},\hat{\rho}_{\omega,t} \}]- \frac{\gamma m k_\text{\tiny{B}}T}{\hbar^2}[\hat{x},[\hat{x},\hat{\rho}_{\omega,t}]]\\
&-\frac{\gamma}{16 m k_{\text{\tiny{B}}}T}[\hat{p},[\hat{p},\hat{\rho}_{\omega,t}]],
\end{split}
\end{equation}
with $\gamma>0$ is a coupling constant, $k_\text{\tiny{B}}$ is Boltzmann's constant and $T$ the temperature. The thermalisation dissipator reads:
\begin{equation}
\begin{split}
\mathcal{D}_{\text{th}}[\hat{\rho}_{\omega,t}]&= \frac{i \Lambda}{4 \hbar}([\hat{p},\{\hat{x},\hat{\rho}_{\omega,t} \}]-[\hat{x},\{\hat{p},\hat{\rho}_{\omega,t} \}])\\
&-\frac{\Lambda (2 \bar{n}+1)}{4 \hbar m \Omega}(m^2 \Omega^2 [\hat{x},[\hat{x},\hat{\rho}_{\omega,t}]]+[\hat{p},[\hat{p},\hat{\rho}_{\omega,t}]]),
\end{split}
\end{equation}
with $\Lambda>0$ a coupling constant and $\bar{n}=(\exp[\frac{\hbar \Omega}{k_\text{\tiny{B}}T}]-1)^{-1}$. Generally speaking, the Caldeira-Leggett dissipator $\mathcal{D}_{\text{CL}}$ models the motion of a Brownian particle weakly coupled to a bath of harmonic oscillators, in the limit of high temperature \cite{breuer2002theory}. The thermalisation dissipator $\mathcal{D}_{\text{th}}$ encodes the interaction with the particles of a bath at temperature $T$ and mean excitation number $\bar{n}$ \cite{Genoni2016}. In the particular scenario of a levitated nanoparticle trapped by a laser in a cold vacuum chamber, the former arises from the interaction between the system and the surrounding residual gas, and the latter from the interaction between the system and the optical modes of the laser \cite{Wu2024}. 

The remaining terms in Eq.~\eqref{levpart} account for the homodyne detection of the particle's position through the scattered photons, as given in Eq.~\eqref{unrav}. The extra factor $K=2\bar{n}_{\text{ph}}+1$ takes into account the fact that the number of thermal photons $\bar{n}_{\text{ph}}$ is not strictly vanishing (see Appendix~\ref{Appbath}), but for practical purposes, one can consider $K \sim 1$ \cite{Setter2018,Wu2024}.

Let us now consider the covariance matrix $\Sigma_t$, given by:
\begin{equation}
\Sigma_t= \left(\begin{array}{cc}
  \Sigma_t(\hat{x})   &  \Sigma_t(\hat{x},\hat{p}) \\
  \Sigma_t(\hat{x},\hat{p})   &  \Sigma_t(\hat{p})
\end{array} \right),
\end{equation}
with $\Sigma_t(\hat{O}_i)=\Sigma_t(\hat{O}_i,\hat{O}_i)$. We restrict the analysis to  Gaussian states, in which case $\Sigma_t$ satisfies a quantum Riccati equation:
\begin{equation}
\frac{\dd \Sigma_t}{\dd t}=\alpha \Sigma_t + \Sigma_t \alpha^T + \delta- \Sigma_t \beta \beta^T \Sigma_t,   
\end{equation}
where $\alpha$, $\beta$ and $\delta$ are the drift, diffusion and backaction matrices, respectively \cite{Wu2024}. Straightforward calculations show that the dynamical evolution for the conditional state $\hat{\rho}_{\omega,t}$ in Eq.~\eqref{levpart} leads to the following expressions for the matrices $\alpha,\beta$ and $\delta$:
\begin{equation} \label{drift}
\alpha=\left(\begin{array}{cc}
  -\frac{1}{2}\Lambda  & \frac{1}{m}  \\
  - m \Omega^2 - 2 \lambda \hbar \xi^{\text{\tiny{R}}}\xi^{\text{\tiny{I}}} & -\frac{1}{2}\Lambda - \gamma
\end{array} \right),     
\end{equation}
\begin{equation} \label{diffback}
\delta= \text{Diag}(d_{11} \quad d_{22}), \qquad \beta= \left(\begin{array}{cc}
   0  &  2 \frac{\sqrt{\lambda}}{\sqrt{K}}\xi^{\text{\tiny{R}}} \\
   0  & 0
\end{array} \right), 
\end{equation}
with $d_{11}=\frac{\hbar^2 \gamma}{8 k_\text{\tiny{B}}mT}+\frac{\hbar \Lambda}{2 m \Omega}(2 \bar{n}+1)$ and $d_{22}= 2 \gamma  m k_\text{\tiny{B}} T + \frac{1}{2}\hbar \Lambda m \Omega (2 \bar{n}+1) + \lambda \hbar^2 K (\xi^{\text{\tiny{R}}})^2$. 
The explicit dependence of the drift, diffusion and back action matrices of the values of the complex parameter $\xi$, shows that   $\Sigma_t$ is unraveling-dependent. Moreover, we stress the fact that the unraveling-dependence is just a consequence of knowing the unraveling used to model the homodyne detection of the position of the particle, as explicitly given by the last line in Eq.~\eqref{levpart}. In an actual experimental setting, when one collects the output signal of the observable of interest, the description of the signal as in Eq.~\eqref{measrec} is equivalent to specifying an unraveling.

The expressions in Eq.~\eqref{drift}  and Eq.~\eqref{diffback} generalize the results in Refs.~\cite{Setter2018,Wu2024}, where only the unraveling corresponding to a standard continuous measurement ($\xi=1$) was considered. In general, the complex phase $\xi$ (between the input state and the local oscillator) in the homodyne detection mechanism can be fixed through a phase plate, as shown in Fig.~\ref{homfig}. This corresponds to performing a measurement of the quadrature $\hat{x}_\xi=(\xi \hat{a}+\xi^* \hat{a}^\dagger)/\sqrt{2}$ (see Appendix \ref{HomApp}). In Fig.~\ref{Unraveling}, we show the time evolution of $\Sigma_t(\hat{x})$ for different values of the parameter $\xi^\text{\tiny{R}}$, each corresponding to a distinct measurement scheme. For $\xi^\text{\tiny{R}} = 1$ (green curve), the indirect homodyne detection is equivalent to a standard continuous measurement of the position. As expected, the position variance decreases over time and eventually stabilizes at a fixed asymptotic value, the lowest among all considered measurement schemes. The localization effect becomes weaker as $\xi^\text{\tiny{R}}$ decreases, as illustrated by the red curve for $\xi^\text{\tiny{R}} = 0.01$. When $\xi^\text{\tiny{R}} = 0$ (orange curve), the position variance still reaches an asymptotic value due to the presence of dissipative dynamics; however, the spread remains significantly larger than in the other two cases.

\begin{figure}
    \centering
    \includegraphics[width=1\linewidth]{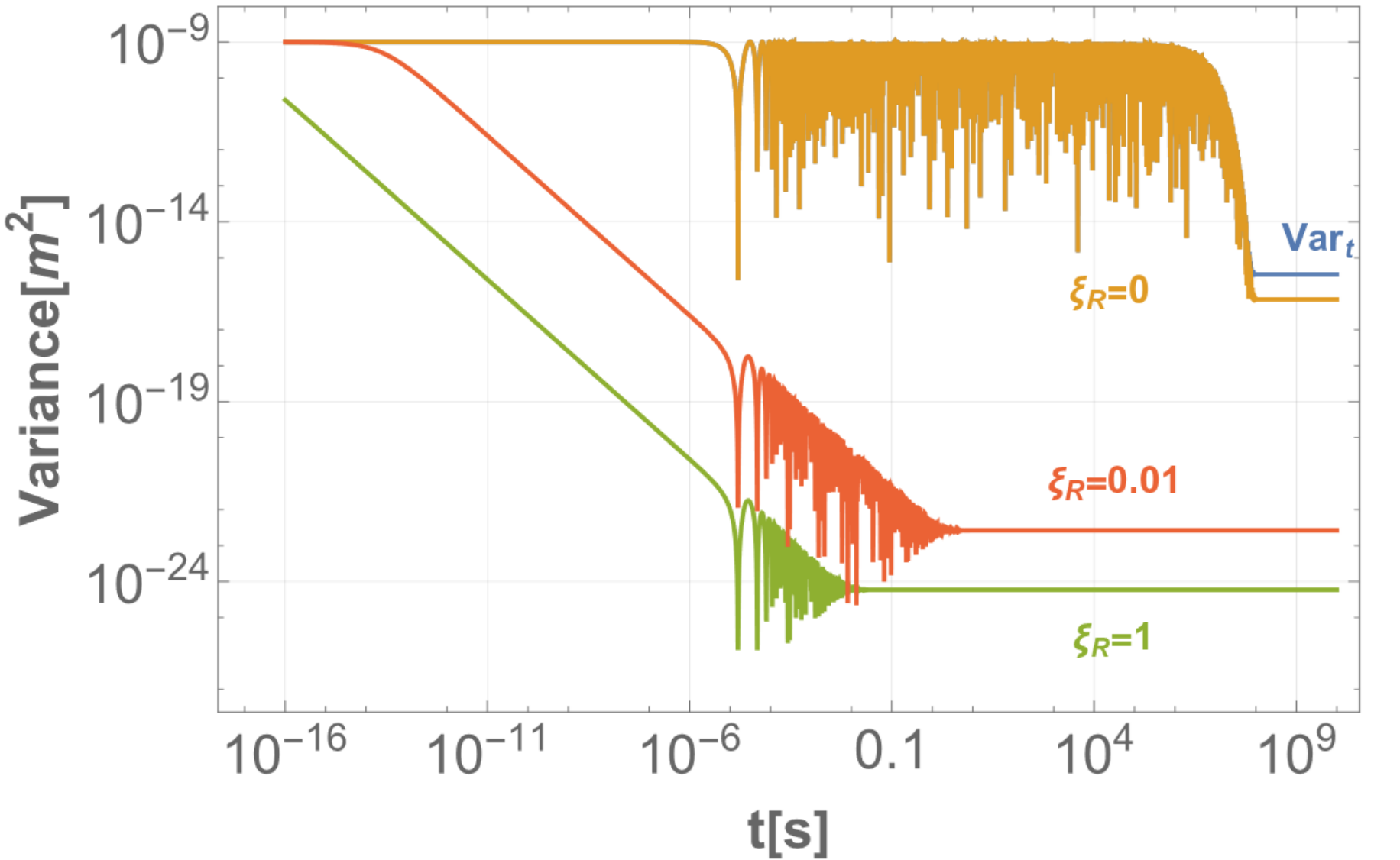}
    \caption{Time evolution of the unraveling-independent variance $\text{Var}_t(\hat{x})$ (blue line), and of the unraveling dependent variance $\Sigma_t(\hat{x})$ for different values of $\xi_\text{\tiny{R}}$. We use the following parameters from \cite{Wu2024}: $m~=~ 10^{-15} \, \mathrm{kg}$, $a_0~=~0.25\times 10^{9} \mathrm{m^{-2}}$, $\lambda~=~10^{26} \, \mathrm{m^{-2}\,Hz}$, $\Omega~=~10^5 \,\mathrm{Hz}$, $\gamma~=~10^{-7}\,\mathrm{Hz}$, $T~=~50 \,\mathrm{K}$, $\Lambda~=~10^{-7}\,\mathrm{Hz}$. The three quantities evolve differently over time.}
    \label{Unraveling}
\end{figure}

For completeness, we include the unraveling-independent covariance matrix $\text{Var}_t$:
\begin{equation}
\text{Var}_t= \left(\begin{array}{cc}
  \text{Var}_t(\hat{x})   &  \text{Var}_t(\hat{x},\hat{p}) \\
  \text{Var}_t(\hat{x},\hat{p})   &  \text{Var}_t(\hat{p})
\end{array} \right),
\end{equation}
where, given two operators $\hat{O}_i$ and $\hat{O}_j$, we have defined $\text{Var}_t(\hat{O}_i,\hat{O}_j)=\mathbb{E}_\omega [\frac{1}{2}\ave{\{\hat{O}_i,\hat{O}_j\}}]-\mathbb{E}[\ave{\hat{O}_i}]\mathbb{E}[\ave{\hat{O}_j}]$, and  $\text{Var}_t(\hat{O}_i) = \text{Var}_t(\hat{O}_i,\hat{O}_i)$. It is easy to show  that $\text{Var}_t$ satisfies as well a quantum Riccati equation, where now:
\begin{equation}
\alpha=\left(\begin{array}{cc}
   -\frac{1}{2} \Lambda  & \frac{1}{m} \\
   - m \Omega^2  & -\frac{1}{2}\Lambda - \gamma
\end{array} \right),
\end{equation}
\begin{equation}
\delta=\text{Diag}(\delta_{11} \quad \delta_{22}), \qquad \beta=0^{2 \times 2},   
\end{equation}
where $\delta_{11}=d_{11}$ and $\delta_{22}=2 \gamma m k_\text{\tiny{B}} T + \frac{1}{2}\hbar\Lambda m \Omega (2 \bar{n}+1)+\lambda \hbar^2 K$. The dependence on $\xi$, therefore on the unraveling, has vanished.
The behavior of $\text{Var}_t(\hat{x})$ is shown in Fig.~\ref{Unraveling}. The overall dynamics resembles the case with $\xi^\text{\tiny{R}}=0$, however the asymptotic value is larger than all the cases already considered. This difference is due to the presence of the additional diffusive term $\lambda \hbar^2 K$, introduced by the homodyne monitoring. 

\section{(In)distinguishability of different unravelings of a master equation} \label{SectBell}

We now turn to the question of whether the nonlinear quantities introduced earlier can be {\it operationally} used to distinguish between different unravelings of the same master equation. From the previous analysis, as well as from the existing literature~\cite{Ghirardi1990b,Pearle1997,Wiseman2001,Adler2007,Caiaffa2017}, it is clear that these quantities differ for different unravelings. The key question is whether this difference can be used to {\it experimentally} discriminate the unravelings.

According to what is described in the previous section and in Ref.~\cite{Piñol2024}, these nonlinear quantities are not  directly inferred from experimental data, but rather computed from the output signal, knowing the unraveling, i.e. the stochastic master equation that models the measurement procedure. For instance, in the example of the previous section, the possibility of computing quantities such as $\Sigma_t(\hat{x})$ arises only after identifying the output signal of the position of the particle with the structure in Eq.~\eqref{measrec}, which has a one-to-one correspondence with the unraveling in the last line of Eq.~\eqref{levpart}. In other words, without specifying the signal (and thus the unraveling), it is not possible to reconstruct quantities such as $\Sigma_t(\hat{x})$.

It is important to remark that nonlinear quantities cannot be directly inferred from experimental data.  Take, for example, \(\langle \hat{O} \rangle_{t}\): by definition, this represents the expectation value of the observable \(\hat{O}\) and, as such, can only be computed by repeating a measurement. However, these repeated measurements should be performed for a {\it fixed} realization of the noise in order to give \(\langle \hat{O} \rangle_{t}\), which is impossible. For the case of the levitated nanoparticle of the previous section, it would imply fixing the realization that defines the dynamics of the conditional density operator in Eq.~\eqref{levpart} for each of the runs of the experiment, which in general cannot be done.

Therefore, one can compute nonlinear quantities only after knowing the unraveling which is under consideration, together with the output signal. The distinction is provided at the source, by choosing one unraveling. In other words, the access to nonlinear quantities which depend on conditional quantum expectation values (even if the noise conditions could be perfectly replicated—which cannot, because that would imply having control of the measurement outcomes) arises only after identifying the measurement record with a specific unraveling, through which the corresponding mathematical computation of the nonlinear quantity of interest can be carried out. There is no possibility of computing the nonlinear quantity directly from the experimental data. As such, the answer to the question raised at the beginning of the section is negative.

The case is different for ${\mathbb E}_\omega[\langle \hat O^2 \rangle_{\omega,t}]$ or $({\mathbb E}_\omega[\langle \hat O \rangle_{\omega,t}])^2$; the first quantity is mathematically equivalent to $\text{Tr}[\hat A^2 \hat{\rho}_t]$, and the second to $(\text{Tr}[\hat A \hat{\rho}_t])^2$, where $\hat{\rho}_t$ is the averaged density matrix. In both cases, the quantum expectation can be  computed directly by performing a measurement of the statistical ensemble represented by $\hat{\rho}_t$, which is supposed to be accessible, without passing through the unraveling that generated it. 

Another, more dramatic, consequence is that, if quantities like ${\mathbb E}_\omega[\langle \hat O \rangle^2_{\omega,t}]$ were directly accessible, one could establish a protocol for  superluminal communication. To show this, let us consider the dynamical evolution of a bipartite system, where one of the parties, let us say, Alice, is continuously monitoring a local observable $\hat{L}=\hat{L}_\text{\tiny{A}} \otimes \hat{1}_\text{\tiny{B}}$, where $\hat{1}_\text{\tiny{B}}$ is the identity operator for the second party, say Bob. He can make local measurements of some observable $\hat{O}=\hat{1}_\text{\tiny{A}} \otimes \hat{O}_\text{\tiny{B}}$. From the evolution of the conditional density operator $\hat{\rho}_{\omega,t}$ in Eq.~\eqref{eq:cme}, and setting for simplicity $\hat{H}=0$, the dynamical evolution of Bob's reduced density operator $\hat{\rho}_{\omega,t}^\text{\tiny{B}}=\text{Tr}_\text{\tiny{A}}[\hat{\rho}_{\omega,t}]$ reads:
\begin{equation} \label{eq:tdjyfhs}
\begin{split}
\dd \hat{\rho}_{\omega,t}^{\text{\tiny{B}}}=2 \sqrt{\lambda}\xi^{\text{\tiny{R}}} \left(\text{Tr}_\text{\tiny{A}}[\hat{L}\hat{\rho}_{\omega,t}]-\ave{\hat{L}}_t\hat{\rho}_{\omega,t}^{\text{\tiny{B}}} \right)\dd W_t.
\end{split}
\end{equation}
As this master equation is purely stochastic, it follows that the expectation value $\mathbb{E}_\omega[\text{Tr}_\text{\tiny{B}}[\hat{O}\hat{\rho}_{\omega,t}^{\text{\tiny{B}}}]]$ is constant over time, as expected because the continuous monitoring of the observable $\hat{L}$ by construction affects only Alice and cannot affect Bob.  It is another way of expressing the no-signaling condition. Moreover, if the initial state is separable, i.e. $\hat{\rho}_{\omega,0}=\hat{\rho}_{\text{\tiny{A}},0} \otimes \hat{\rho}_{\text{\tiny{B}},0}$,  from the above equation it follows the stronger result $\dd \hat{\rho}_{\omega,t}^{\text{\tiny{B}}} = 0$: now, the state of Bob does not change, not only in the average, but also for each realization. This is again entirely expected: if the states are initially factorized and the dynamics---included the measurements---is also factorized, Alice and Bob are completely disconnected from each other.

However, if Alice and Bob share pairs of particles in an initially entangled state, and Bob can measure quantities which are nonlinear in $\hat{\rho}_{\omega,t}^{\text{\tiny{B}}}$, without previously knowing what Alice will do on her side, superluminal communication could be established between them. As a matter of fact, let us consider that Alice and Bob share pairs of $1/2$-spin particles, let us say in the singlet state:
\begin{equation} \label{singlet}
\ket{s}=\frac{1}{\sqrt{2}}(\ket{\uparrow}\otimes \ket{\downarrow}-\ket{\downarrow} \otimes \ket{\uparrow}). 
\end{equation}
In particular, let Alice continuously monitor the $n'$-component of the spin of her particles (where the components $n'$ and $n$ do not necessarily coincide), thus $\hat{L}=\hat{\sigma}_{n'} \otimes \hat{1}_\text{\tiny{B}}$. On the other side, Bob is interested in the dynamical evolution of the covariance $\Sigma_t(\hat{\sigma}_n)~=~1~-~(\text{Tr}_\text{\tiny{B}}[\hat{\sigma}_n \hat{\rho}_{t,\omega}^{\text{\tiny{B}}}])^2$ (we have used the fact that $\hat{\sigma}_n^2=\hat{1}_\text{\tiny{B}}$) of the $n$-component of the spin of his particles. Starting from Eq.~\eqref{eq:tdjyfhs}, straightforward calculations show that: 
\begin{equation}
\begin{split}
\dd \Sigma_t(\hat{\sigma}_n)\!\!=\!\!-\!4 \lambda (\xi^{\text{\tiny{R}}})^2 S_t^2(n,n') \dd t \!-\!4 \sqrt{\lambda}\xi^{\text{\tiny{R}}}\ave{\hat{\sigma}_n}_t S_t(n,n')\dd W_t,
\end{split} \label{Sigmatdyn}
\end{equation}
with $S_t(n,n')~\!\!\!=\!\!\!~\text{Tr}[\hat{\sigma}_{n'}~ \otimes~ \hat{\sigma}_{n} \hat{\rho}_{\omega,t}]~\!\!-\!\!~\text{Tr}_\text{\tiny{A}}[\hat{\sigma}_{n'} \hat{\rho}_{\omega,t}^{\text{\tiny{A}}}]\text{Tr}_\text{\tiny{B}}[\hat{\sigma}_n \hat{\rho}_{\omega,t}^{\text{\tiny{B}}}]$. 
From the family of unravelings in Eq.~\eqref{unrav}, we consider the  cases $\xi=i$, and $\xi=1$. For the first case, we have:
\begin{equation} \label{Sigmai}
\dd \mathbb{E}_\omega[ \Sigma_{t}^{\xi=i}(\hat{\sigma}_n)] = 0,    
\end{equation}
i.e. the variance does not change in the average, while in the second case it does because:
\begin{equation} \label{Sigma1}
\dd \mathbb{E}_\omega [\Sigma_t^{\xi=1}(\hat{\sigma}_n)]=- 4 \lambda  \mathbb{E}_\omega[S_{t}^2(n,n')] \dd t,    
\end{equation}
with at least $S_0^2(n,n')=(n \cdot n')^2$ in general different from 0.

\begin{figure}[t!] 
    \centering
     \includegraphics[width=1\linewidth]{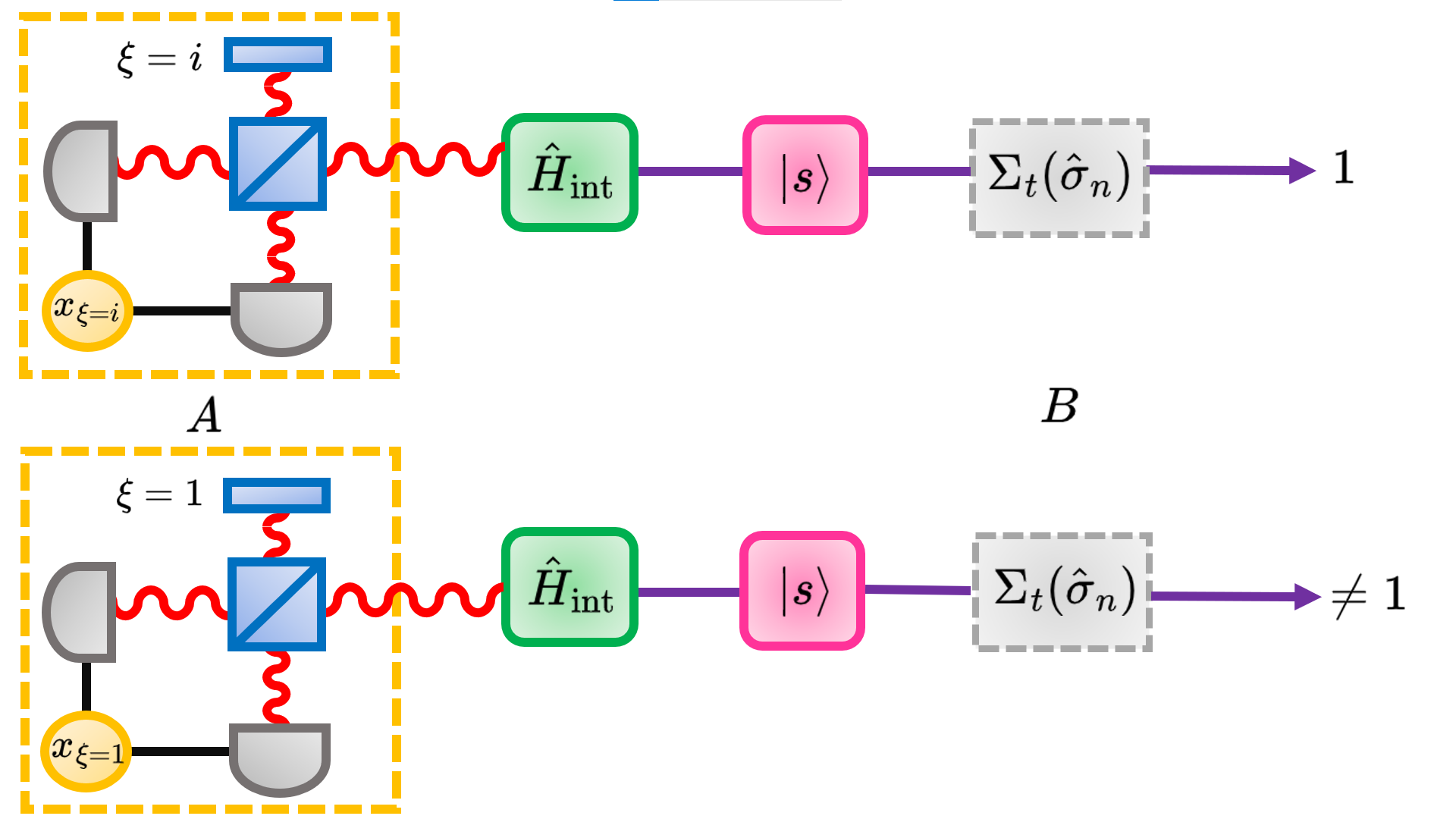}
    \caption{Superluminal protocol established between two parties, Alice (A) and Bob (B), if $\Sigma_t(\hat{\sigma}_n)$ were accessible. If they share pairs of entangled particles in the state $\ket{s}$ [cf. Eq.~\eqref{singlet}], and Alice continuously monitors the spin of her particles, Bob would be in a position to know with certainty Alice's choice regarding the phase $\xi$ with which she performs homodyne detection after coupling her particles with light. It would suffice for him to determine $\Sigma_t(\hat{\sigma}_n)$ on his particles. If Alice chooses $\xi=i$, then Bob would see the outcome $\Sigma_t(\hat{\sigma}_n)=1$. In contrast, if she chooses $\xi=1$, Bob would see the outcome $\Sigma_t(\hat{\sigma}_n)\neq 1$. This is independent of the spatial distance separating the two parties, opening the way to superluminal signaling.}
    \label{sigfig}
\end{figure}

If $\Sigma_t(\hat{\sigma}_n)$ were directly accessible for Bob, i.e. without Alice having previously told him which kind of measurement she would  perform, he would obtain $\mathbb{E}_\omega[\Sigma_t(\hat{\sigma}_n)]= \mathbb{E}_\omega[\Sigma_0(\hat{\sigma}_n)] = 1$ if Alice continuously monitors the spin of her particles by using the unraveling in Eq.~\eqref{unrav} with $\xi=i$, and $\mathbb{E}[\Sigma_t(\hat{\sigma}_n)]\neq 1$ for at least some time, if she continuously monitors her particles using the unraveling corresponding to $\xi=1$. A schematic representation of this protocol is shown in Fig.~\ref{sigfig}. Bob would then be able to determine Alice's choice from a distance, thus allowing superluminal communication between them. 

However, as discussed earlier, nonlinear quantities are not  accessible without knowing the unraveling and, as such, they cannot be used to distinguish between different unravelings; therefore, the superluminal protocol cannot be established.

The situation changes if Alice informs Bob of the measurement she performs and shares the measurement signal; in this case, Bob can reconstruct \({\mathbb E}_\omega [\Sigma(\hat{\sigma}_n)]\). This does not violate relativity, as Alice and Bob must now establish a classical communication protocol. Moreover, Bob does not gain any additional knowledge on the type of measurement Alice performed beyond what is already contained in the measurement procedure disclosed by her.

Finally, we remark that, as already known, a measurement of a quantity linear in the density operator does not lead to superluminal signaling. In the above example, let us consider for concreteness that Bob decides to measure $\hat{\sigma}_n$. Therefore, the dynamical evolution of $\ave{\hat{\sigma}_n}_t=\text{Tr}_\text{\tiny{B}}[\hat{\sigma}_n \hat{\rho}_{t,\omega}^\text{\tiny{B}}]$ reads
\begin{equation}
\dd \ave{\hat{\sigma}_n}_t=2 \sqrt{\lambda}\xi^{\text{\tiny{R}}} S_t(n,n') \dd W_t,    
\end{equation}
from which it immediately follows that, in analogy to Eqs.~\eqref{Sigmai} and \eqref{Sigma1}
\begin{equation}
\dd \mathbb{E}_\omega[\ave{\hat{\sigma}_n}_t^{\xi=i}]=0=\dd \mathbb{E}_\omega[\ave{\hat{\sigma}_n}_t^{\xi=1}],    
\end{equation}
and thus, Bob cannot distinguish the type of measurement that Alice performed.

\section{Conclusion}

There exist several stochastic unravelings for a given master equation. In light of this, a relevant question—from both a foundational and a practical perspective—is whether these unravelings can be distinguished. This issue can be addressed at both the mathematical and operational levels.

From a mathematical standpoint, it is well known that different unravelings evolve the wave function differently, and as such that can be distinguished~\cite{Ghirardi1990b,Pearle1997,Wiseman2001,Adler2007,Caiaffa2017}, not least because the very notion of different unravelings implies some form of differentiation. Thus, the real question is whether they can be distinguished operationally.

As we have discussed, the answer is negative. Physical quantities are either unraveling-independent, such as \(\mathbb{E}_\omega [\langle \hat{O}^n \rangle_t]\), and thus cannot be used to discriminate between unravelings; or they are unraveling-dependent, such as \(\mathbb{E}_\omega [\langle \hat{O} \rangle^n_t]\) for \(n > 1\), but they are not operationally accessible unless the unraveling, or measurement procedure, which nonlinear quantities are supposed to distinguish, is already known.

\section{Acknowledgments}
The authors wish to thank D. Keys and M. Paternostro for useful comments on the first draft of the paper. They acknowledge support from the EU EIC Pathfinder project QuCoM (101046973), INFN and the University of Trieste. A.B. acknowledges further support from  the PNRR MUR projects PE0000023-NQSTI. 

\appendix

\section{A quick review on homodyne detection} \label{HomApp}

We will follow the derivation presented in Refs.~\cite{Albarelli2024,Wiseman1994} to review the most relevant aspects of homodyne detection. We take a system $S$, of which we are interested in continuously monitoring the operator $\hat{L}$. Let us consider the interaction between the system $S$ and the environment $A$, where the latter is modeled as a continuum of bosonic modes, with creation and annihilation operators $\hat{a}^\dagger$ and $\hat{a}$, respectively (thus, they satisfy $[\hat{a},\hat{a}^\dagger]=1$). If the total Hamiltonian is $\hat{H}_{\text{tot}}=\hat{H}_S+\hat{H}_A + \hat{H}_{\text{int}}$, where $\hat{H}_S$ acts on $S$, $\hat{H}_A$ on $A$ and $\hat{H}_\text{int}$ encodes the interaction between system and environment. In what follows, we will work in the interaction picture with respect to the Hamiltonian $\hat{H}=\hat{H}_S+\hat{H}_A$. Let us set the interaction Hamiltonian to be given by
\begin{equation}
\hat{H}_{\text{int}}=i \hbar \sqrt{\frac{\lambda}{\dd t}} \left(\hat{L} \otimes \hat{a}^\dagger - \hat{L}^\dagger \otimes \hat{a} \right),     
\end{equation}
where $\lambda$ is a suitable coupling constant. If the total state of system and environment at a time $t$ is $\ket{\psi_t} \otimes \ket{0}$, first we let it evolve under the unitary evolution due to $\hat{H}_\text{int}$. Then, the environment is projected on the eigenstates $\{\ket{x_\xi}\}$ of the operator $\hat{x}_\xi= (\xi \hat{a}+\xi^* \hat{a}^\dagger)/\sqrt{2}$, where $\xi$ is a complex number satisfying $|\xi|^2=1$. Thus, at a time $t+\dd t$, we have that the unnormalized state of the system reads
\begin{equation}
\ket{\widetilde{\psi}_{t+\dd t}}=\left( \hat{1} \otimes \bra{x_\xi} \right)e^{-\frac{i}{\hbar}\hat{H}_{\text{int}}\dd t}\left(\ket{\psi_{t}} \otimes \ket{0} \right).   
\end{equation}
Straightforward calculations show that
\begin{equation}
\begin{split}
\ket{\widetilde{\psi}_{t+\dd t}}&=\bra{x_\xi}0 \rangle \left(\hat{1}+\sqrt{\lambda \dd t}\frac{\bra{x_\xi}1 \rangle}{\bra{x_\xi}0 \rangle}\hat{L} \right.\\
&-\left.\frac{1}{2}\lambda \dd t \left(\hat{L}^\dagger \hat{L} -\sqrt{2}\frac{\bra{x_\xi}2\rangle}{\bra{x_\xi}0\rangle} \hat{L}^2 \right)  \right)\ket{\psi_t}+\mathcal{O}(\dd t)^{3/2}.    
\end{split}
\end{equation}
From the eigenvalue equation $\hat{x}_\xi \ket{x_\xi}=x_\xi \ket{x_\xi}$, the following identities hold
\begin{equation}
\bra{x_\xi}1\rangle=\sqrt{2}\xi x_\xi \bra{x_\xi}0 \rangle, \qquad \bra{x_\xi}2 \rangle=\frac{\xi^2}{\sqrt{2}}(2 x_\xi^2-1)\bra{x_\xi}0 \rangle,    
\end{equation}
so that at time $t+\dd t$ we have
\begin{equation}
 \begin{split}
\ket{\widetilde{\psi}_{t+\dd t}}&=\bra{x_\xi}0\rangle \left(\hat{1}+\sqrt{2 \lambda \dd t}\xi x_\xi \hat{L} \right.\\
&\left.-\frac{1}{2}\lambda \dd t \left(\hat{L}^\dagger \hat{L}-\xi^2 (2 x_\xi^2-1)\hat{L}^2 \right) \right)\ket{\psi_t}+\mathcal{O}(\dd t)^{3/2}.
\end{split}   
\end{equation}
Thus, the probability of obtaining the outcome $x_\xi$ after projection on the environment of the total state at time $t+\dd t$ reads  
\begin{equation}
\begin{split}
&p(x_\xi)_{t+\dd t}=\norm{\widetilde{\psi}_{t+\dd t}}^2\\
&=|\bra{x_\xi}0\rangle|^2 \left(1 + \sqrt{2 \lambda \dd t}x_\xi \ave{\xi \hat{L}+\xi^* \hat{L}^\dagger}_t \right.\\
&\left.+ \frac{1}{2}(2 x_\xi^2-1)\lambda \dd t (2 \ave{\hat{L}^\dagger \hat{L}}_t+ \ave{\xi^2 \hat{L}^2 + (\xi^*)^2 (\hat{L}^\dagger)^2}_t) \right)\\
&+\mathcal{O}(\dd t)^{3/2}.
\end{split}
\end{equation}
Due to the vacuum fluctuations, the probability distribution for $x_\xi$ at a time $t$ is given by $p(x_\xi)_t=|\bra{x_\xi}0 \rangle|^2=(1/\sqrt{\pi})e^{-x_\xi^2}$ \cite{Albarelli2024}. Therefore, up to order $\sqrt{\dd t}$, we can approximate $p(x_\xi)_{t+\dd t}$ as 
\begin{equation}
p(x_\xi)_{t+\dd t} = \frac{1}{\sqrt{\pi}} \exp \left[-\left(x_\xi - \sqrt{\frac{\lambda \dd t}{2}}\ave{\xi \hat{L}+\xi^* \hat{L}^\dagger}_t \right)^2 \right].    
\end{equation}
This allows us to identify the random variable $x_\xi$ as a stochastic random variable satisfying
\begin{equation} \label{xxi}
x_\xi=\sqrt{\frac{\lambda \dd t}{2}}\ave{\xi \hat{L}+\xi^* \hat{L}^\dagger}_t+\frac{\dd W_t}{\sqrt{2 \dd t}},    
\end{equation}
where $W_t$ is a Wiener process. Under this approximation, the above expressions simplify to
\begin{equation}
\begin{split}
\ket{\widetilde{\psi}_{t+\dd t}}&=\bra{x_\xi}0\rangle \left(\hat{1}+ \sqrt{2 \lambda \dd t}\xi x_\xi \hat{L}-\frac{1}{2}\lambda \dd t \hat{L}^\dagger \hat{L} \right)\ket{\psi_t}\\
&+ \mathcal{O}(\dd t)^{3/2},    
\end{split}
\end{equation}
\begin{equation}
p(x_\xi)_{t+\dd t}\!=\!|\!\bra{x_\xi}0 \rangle\!|^2\!\left(1 \!+\!\sqrt{2\lambda \dd t}x_\xi \ave{\xi \hat{L}\!+\! \xi^* \hat{L}^\dagger}_t\right)+\mathcal{O}(\dd t)^{3/2}.
\end{equation}
Therefore, the normalized state at a time $t+\dd t$ is given by
\begin{equation}
\ket{\psi_{t+\dd t}}=\frac{\ket{\widetilde{\psi}_{t+\dd t}}}{\sqrt{p(x_\xi)_{t+\dd t}}}.    
\end{equation}
Expanding up to order $\dd t$, and after dropping out a global phase due to $\bra{x_\xi}0\rangle$ we obtain
\begin{equation}
\begin{split}
\dd \ket{\psi_t}&\!=\!  \left(-\frac{\lambda}{2}\left( \!\hat{L}^\dagger \hat{L}\!-\!\ave{\xi \hat{L}\! +\! \xi^* \hat{L}^\dagger}_t \xi \hat{L}\!+\!\frac{1}{4} \ave{\xi \hat{L} + \xi^* \hat{L}^\dagger}_t^2 \right)\dd t \right.\\
&\left.+ \sqrt{\lambda}\left(\xi \hat{L} -\frac{1}{2}\ave{\xi \hat{L} + \xi^* \hat{L}^\dagger}_t \right)\dd W_t \right)\ket{\psi_t},    
\end{split}
\end{equation}
and if the operator $\hat{L}$ is Hermitian, the above equation simplifies to 
\begin{equation}
\begin{split}
\dd \ket{\psi_t}&\!=\!\left(\!-\!\frac{\lambda}{2}(|\xi|^2 \hat{L}^2-2 \xi \xi^\text{\tiny{R}} \hat{L}\ave{\hat{L}}_t+(\xi^\text{\tiny{R}})^2 \ave{\hat{L}}_t^2)\dd t \right.\\
&\left.+\sqrt{\lambda} (\xi \hat{L} - \xi^\text{\tiny{R}} \ave{\hat{L}}_t)\dd W_t\right)\ket{\psi_t},
\end{split}
\end{equation}
thus recovering the result of Eq.~\eqref{unrav} once one switches back to the Schr\"odinger picture. As a final remark, the measurement record $y_t$ satisfies the following differential equation [cf. Eq.~\eqref{xxi}]:
\begin{equation}
\dd y_t=\sqrt{\frac{\dd t}{2\lambda}} x_\xi=\frac{1}{2}\ave{\xi \hat{L}+\xi^* \hat{L}^\dagger}_t \dd t+\frac{1}{2 \sqrt{\lambda}} \dd W_t.
\end{equation}

\section{Continuous monitoring on a finite temperature bath} \label{Appbath}

We generalize the results presented in Refs. \cite{Wiseman1994,Albarelli2024}, and consider the continuous monitoring of a system surrounded by an environment with finite temperature. It is more convenient to work at the level of the Wigner function for the thermal bath \cite{Gardiner2000}. Therefore, the initial state of the system and bath can be represented as 
\begin{equation}
W_t(\alpha,\alpha^*)=\frac{2}{\pi K}\exp \left[-\frac{2}{K}\alpha \alpha^* \right]\hat{\rho}_t,   
\end{equation}
with $K^{-1}=\tanh(\frac{\hbar \Omega}{2 k_\text{\tiny{B}}T})$ and where $\alpha$ is the eigenvalue of the coherent state $\ket{\alpha}$, i.e., $\hat{a}\ket{\alpha}=\alpha \ket{\alpha}$. Once again, we let this initial state evolve under the Hamiltonian $\hat{H}_{\text{int}}$, so that 
\begin{equation} \label{Wignerexp}
W_{t+\dd t}(\alpha,\alpha^*)=e^{-\frac{i}{\hbar}\hat{H}_\text{int}\dd t}W_t(\alpha,\alpha^*)e^{\frac{i}{\hbar}\hat{H}_\text{int}\dd t},
\end{equation}
where we can use the following results \cite{Gardiner2000} when applying the unitary evolution operators on the initial state of the system and bath
\begin{equation}
\begin{split}
\hat{a}W_t(\alpha,\alpha^*)&= \left(\alpha + \frac{1}{2}\frac{\partial}{\partial \alpha^*} \right)W_t(\alpha,\alpha^*),\\
\hat{a}^\dagger W_t(\alpha,\alpha^*)&= \left(\alpha^*-\frac{1}{2}\frac{\partial}{\partial \alpha} \right)W_t(\alpha,\alpha^*),\\
W_t(\alpha,\alpha^*)\hat{a} &= \left(\alpha-\frac{1}{2}\frac{\partial}{\partial {\alpha^*}} \right)W_t(\alpha,\alpha^*),\\
W_t(\alpha,\alpha^*)\hat{a}^\dagger &= \left(\alpha^*+\frac{1}{2}\frac{\partial}{\partial \alpha} \right)W_t(\alpha,\alpha^*).
\end{split}
\end{equation}
As in the previous case, we are interested in the measurement of the observable $\hat{x}_\xi=(\xi \hat{a}+\xi^* \hat{a}^\dagger)/\sqrt{2}$. Therefore, let us define 
\begin{equation}
x_\xi=\frac{\xi \alpha + \xi^* \alpha^*}{\sqrt{2}}, \qquad y_\xi=-i\frac{(\xi \alpha- \xi^* \alpha^*)}{\sqrt{2}},   
\end{equation}
and express the resulting state of the system $W_{t+\dd t}(\alpha,\alpha^*)$ in terms of $x_\xi$ and $y_\xi$. The conditional state $W^{(c)}_{t+\dd t}(x_\xi)$ is obtained by integrating the evolved state $W_{t+\dd t}(x_\xi,y_\xi)$ over all possible values of $y_\xi$. Straightforward calculations show that, to order $\sqrt{\dd t}$, we obtain 
\begin{equation}
\begin{split}
&{W}^{(c)}_{t+\dd t}(x_\xi)\! \propto\! G(x_\xi,0,K/2)\!\left(\!\hat{\rho}_t \!+\! \sqrt{\lambda \dd t}\frac{\sqrt{2}}{K}x_\xi \!\left(\!(N\!+\!1)\xi \hat{L}\hat{\rho}_t \right.\right.\\
&\left.\left.- N \xi^* \hat{L}^\dagger \hat{\rho}_t -N \xi \hat{\rho}_t \hat{L} + (N+1) \xi^*\hat{\rho}_t\hat{L}^\dagger \right) \right),
\end{split}
\end{equation}
where $G(x_\xi,0,K/2)$ is a Gaussian distribution with mean zero and variance $K/2$. From this result, the probability of obtaining the outcome $x_\xi$, is given by
\begin{equation}
\begin{split}
p(x_\xi)_{t+\dd t}&=\text{Tr}[W_c(t+\dd t)]\\
&\propto G(x_\xi,0,K/2)\left(1+\sqrt{\lambda \dd t}\frac{\sqrt{2}}{K}x_\xi \ave{\xi \hat{L}+\xi^* \hat{L}^\dagger} \right)\\
&\approx G \left(x_\xi,\sqrt{\frac{\lambda \dd t}{2}}\ave{\xi \hat{L}+\xi^* \hat{L}^\dagger},K/2 \right),
\end{split}
\end{equation}
where the last approximation is valid to order $\sqrt{\dd t}$. Therefore, we can equivalently set $x_\xi$ to be the following stochastic random variable 
\begin{equation} \label{xxiWigner}
x_\xi=\sqrt{\frac{\lambda \dd t}{2}}\ave{\xi \hat{L}+\xi^* \hat{L}^\dagger}_t + \sqrt{K}\frac{\dd W_t}{\sqrt{2 \dd t}},    
\end{equation}
where $\dd W_t$ is a Wiener process. Going to the next order in the expansion of $W_{t+\dd t}^{(c)}(x_\xi)$ in Eq.~\eqref{Wignerexp}, there will be terms proportional to $(x_\xi^2-K/2)$, which will not contribute to the total dynamics, according to Eq.~\eqref{xxiWigner}. In contrast, terms proportional to $(x_\xi^2+K/2)$ will contribute with to the total dynamics of the system, with a factor of $K$. Therefore, the state $\hat{\rho}_{t+\dd t}$ reads
\begin{equation}
\begin{split}
\hat{\rho}_{t+\dd t}&=\frac{W^{(c)}_{t+\dd t}(x_\xi)}{p(x_\xi)_{t+\dd t}}\\
&=\hat{\rho}_t - \frac{\sqrt{\lambda}}{\sqrt{K}} \mathcal{H}[(N+1)\xi \hat{L}-N \xi^* \hat{L}^\dagger] \hat{\rho}_t \dd W_t \\
&+ \lambda \left((N+1) \mathcal{D}[\hat{L}]\hat{\rho}_t + N \mathcal{D}[\hat{L}^\dagger]\hat{\rho}_t \right)\dd t,
\end{split}
\end{equation}
where $\mathcal{D}[\hat{O}]\hat{\rho}_t$ and $\mathcal{H}[\hat{O}]\hat{\rho}_t$ are super-operators acting on $\hat{\rho}_t$ as
\begin{equation}
\begin{split}
\mathcal{D}[\hat{O}]\hat{\rho}
_t &=\hat{O}\hat{\rho}_t \hat{O}^\dagger - \frac{1}{2}\{\hat{O}^\dagger \hat{O},\hat{\rho}_t \},\\
\mathcal{H}[\hat{O}]\hat{\rho}_t
&=\hat{O}\hat{\rho}_t+\hat{\rho}_t \hat{O}^\dagger - \ave{\hat{O}+\hat{O}^\dagger}_t \hat{\rho}_t.
\end{split}
\end{equation}
Thus, we obtain that the conditional master equation reads
\begin{equation}
\begin{split}
\dd \hat{\rho}_t &=\lambda \left((N+1)\mathcal{D}[\hat{L}]\hat{\rho}_t + N \mathcal{D}[\hat{L}^\dagger]\hat{\rho}_t \right)\dd t\\
&-\frac{\sqrt{\lambda}}{\sqrt{K}} \mathcal{H}[(N+1)\xi \hat{L}-N \xi^* \hat{L}^\dagger]\hat{\rho}_t \dd W_t.
\end{split}
\end{equation}

\section{Some examples of unraveling-dependent quantities} \label{AppExamples}

In what follows, we will discuss some examples of unraveling-dependent quantities. From the family of unravelings in Eq.~\eqref{unrav}, and for illustrative purposes, we will consider two extreme cases: $\xi = \xi^\text{\tiny{R}}$ (i.e., $\xi^\text{\tiny{I}} =0$), corresponding to a standard continuous quantum measurement; and $\xi =-i$ (i.e., $\xi^\text{\tiny{R}} =0$), corresponding to a linear dynamics with the addition of a stochastic potential.

The first choice arises in the context of Gaussian continuous measurements \cite{jacobs2014quantum, wiseman2009quantum}, as well as dynamical collapse models \cite{Bassi2003,Bassi2013}. The dynamics is nonlinear in the wavefunction, but norm-preserving:
\begin{equation} \label{contmeas}
\dd \ket{\psi^{\text{\tiny A}}_t}=\left(-\frac{i}{\hbar}\hat{H}\dd t - \frac{\lambda}{2}(\hat{L}^{\text{\tiny A}}_t)^2 \dd t+ \sqrt{\lambda}\hat{L}^{\text{\tiny A}}_t\dd W_t \right)\ket{\psi^{\text{\tiny A}}_t}, 
\end{equation}
where $\hat{L}^{\text{\tiny A}}_t \equiv \hat{L}-\ave{\hat{L}}^{\text{\tiny A}}_t$. 

The second and third terms {induce the localization} of the wave function towards {an eigenstate} of the operator $\hat{L}$, which occurs randomly in accordance with the Born probability rule.

{The second choice, in contrast to the previous one, leads to a linear dynamics}:
\begin{equation} \label{unitary}
\dd \ket{\psi^{\text{\tiny B}}_t}=\left(-\frac{i}{\hbar}\hat{H}\dd t - \frac{\lambda}{2}\hat{L}^2 \dd t - i\sqrt{\lambda}\hat{L}\dd W_t \right)\ket{\psi^{\text{\tiny B}}_t}.
\end{equation}
The second term, proportional to $\hat{L}^2$, is  the It\^o correction to the otherwise unitary evolution {encoded both in the deterministic contribution of the Hamiltonian $\hat{H}$, as well as the stochastic term proportional to $\hat{L}$}, and disappears from the solution of the equation. This unraveling is also norm-preserving; however, it does not lead to the collapse of the wave function.

As mentioned in the Introduction, one may distinguish between different unravelings through the examination of quantities that cannot be expressed in terms of the density operator $\hat{\rho}_t$. An example of this is $\ave{\hat{O}}^2_t$. For the unraveling in Eq.~\eqref{contmeas} we have: 
\begin{equation} \label{cmav2}
\begin{split}
\dd {\ave{\hat{O}}_{t}^{\text{\tiny A}}}^2&= 2 \ave{\hat{O}}_{t}^{\text{\tiny A}} \left(-\frac{i}{\hbar}\ave{[\hat{O},\hat{H}]}_{t}^{\text{\tiny A}} - \frac{\lambda}{2}\ave{[\hat{L},[\hat{L},\hat{O}]]}_{t}^{\text{\tiny A}}  \right)\dd t\\
&+\lambda (\ave{\{\hat{O},\hat{L} \}}_{t}^{\text{\tiny A}} - 2 \ave{\hat{O}}_{t}^{\text{\tiny A}} \ave{\hat{L}}_{t}^{\text{\tiny A}})^2\dd t\\
&+2 \ave{\hat{O}}_{t}^{\text{\tiny A}} \sqrt{\lambda}(\ave{ \{\hat{O},\hat{L}\}}_{t}^{\text{\tiny A}}-2 \ave{\hat{O}}_{t}^{\text{\tiny A}} \ave{\hat{L}}_{t}^{\text{\tiny A}})\dd W_t,
\end{split}
\end{equation}
whereas the unraveling in Eq.~\eqref{unitary} leads to the following equation:
\begin{equation} \label{unitav2}
\begin{split}
\dd {\ave{\hat{O}}_{t}^{\text{\tiny B}}}^2&=2 \ave{\hat{O}}_{t}^{\text{\tiny B}}\left(-\frac{i}{\hbar}\ave{[\hat{O},\hat{H}]}_{t}^{\text{\tiny B}} \dd t - \frac{\lambda}{2}\ave{[\hat{L},[\hat{L},\hat{O}]]}_{t}^{\text{\tiny B}} \right)\dd t\\
&- \lambda{\ave{[\hat{L},\hat{O}]}_{t}^{\text{\tiny B}}}^2 \dd t+2i \ave{\hat{O}}_{t}^{\text{\tiny B}}\sqrt{\lambda}\ave{[\hat{L},\hat{O}]}_{t}^{\text{\tiny B}}\dd W_t.
\end{split}
\end{equation}

In taking the average over the noise, due to the properties of the Wiener process, the terms proportional to $\dd W_t$  vanish. What remains is different in the two cases, as we will explicitly show with the following two examples.

\subsection{Free-particle}

As a first example, we consider a free particle of mass $m$ undergoing decoherence in position; the simplest choice for Eq.~\eqref{mastereq} is to take
\begin{equation}\label{freepartop}
\hat{H}=\frac{\hat{p}^2}{2m}, \qquad \hat{L}=\hat{x}.
\end{equation}
We will  be interested, for each of the two unravelings, in  the time evolution of the spread in position: $\mathbb{E}_\omega[\Sigma_t(\hat{x})]$ with $ \Sigma_t(\hat{x})~=~\ave{\hat{x}^2}_t-\ave{\hat{x}}^2_t$. 
Let us restrict the analysis to Gaussian states:
\begin{equation} \label{Gaussian}
{\psi}_t(x)=\exp \left[-a_t(x-\overline{x}_t)^2+ i \overline{k}_t x+ \gamma_t   \right].
\end{equation}
In what follows, we will show that the above ansatz is a solution for both unravelings, where $a_t$ and $\gamma_t$ are complex functions of time, whereas $\overline{x}_t$ and $\overline{k}_t$ are real.

Given the Gaussian ansatz in Eq.~\eqref{Gaussian}, one can write the quantum expectation values of observables of interest in terms of the Gaussian parameters as follows:

\begin{equation} \label{Gaussianexps}
\ave{\hat{x}}_t=\overline{x}_t, \qquad \ave{\hat{p}}_t=\hbar \overline{k}_t, \qquad \ave{\hat{x}^2}_t=\frac{1}{4 a_t^{\text{\tiny{R}}}}+ \overline{x}_t^2,
\end{equation}
where $a_t^{\text{\tiny{R}}}$ is the real part of $a_t$. We will use these expressions to derive the dynamical evolution of  $\ave{\hat{x}}^2_t$ for both unravelings and show that it is an unraveling-dependent quantity.
 
\subsubsection{Nonlinear unraveling}

Let us first consider the nonlinear unraveling of  Eq.~\eqref{cmav2}, and the Gaussian ansatz of Eq.~\eqref{Gaussian}. Using the results of Eq.~\eqref{Gaussianexps}, the dynamical evolution of ${\ave{\hat{x}}_{t}^{\text{\tiny{A}}}}^2$ is given by
\begin{equation} \label{x2NLdef}
\dd {\ave{\hat{x}}_{t}^{\text{\tiny{A}}}}^2= \frac{2 \hbar}{m}\overline{x}_t^{\text{\tiny{A}}}\overline{k}_t^{\text{\tiny{A}}} \dd t +\sqrt{\lambda}  \overline{x}_t^{\text{\tiny{A}}} \frac{1}{a_t^{\text{\tiny{A}},\text{\tiny{R}}}} \dd  W_t + \lambda \frac{1}{4 {a_t^{\text{\tiny{A}},\text{\tiny{R}}}}^2}\dd t,
\end{equation}

Thus, we need to determine the evolution of the Gaussian parameters $\overline{x}_t^{\text{\tiny{A}}}$, $\overline{k}_t^{\text{\tiny{A}}}$, and $a_t^{{{\text{\tiny{A}}},\text{\tiny{R}}}}$, and we can do so by following the approach in Ref.~\cite{Bassi2005}. With the choices for $\hat{H}$ and $\hat{O}$ of Eq.~\eqref{freepartop}, the dynamical evolution in the position representation of the nonlinear unraveling in Eq.~\eqref{contmeas} reads
\begin{equation}
\begin{split}
{\dd {\psi_t^{\text{\tiny{A}}}}(x)}&=\left(\frac{i \hbar}{2m}\frac{\partial^2}{\partial x^2}\dd t - \frac{\lambda}{2} (x-\overline{x}_t^{\text{\tiny{A}}})^2\dd t \right.\\
&\left.+\sqrt{\lambda} (\hat{x}-\overline{x}_t^{\text{\tiny{A}}}) {\dd W_t}\right){\psi_t^{\text{\tiny{A}}}}(x),
\end{split}
\end{equation}
where we used Eq.~\eqref{Gaussianexps}. Recognizing that Gaussian states remain Gaussian, and after matching the terms on both sides of the above equation in powers of $x$, we obtain the following equations for the relevant parameters of the Gaussian state:
\begin{equation} \label{spreadcontmeas}
\begin{split}
{\dd a_t^{\text{\tiny{A}}}} &=\left(\lambda -\frac{2i \hbar}{m}{a_t^{\text{\tiny{A}}}}^2 \right)\dd t,\\
{\dd \overline{x}_t^{\text{\tiny{A}}}} &=\frac{\hbar}{m}\overline{k}_t^{\text{\tiny{A}}}\dd t+ \frac{\sqrt{\lambda} }{2 a_t^{\text{\tiny{A}},\text{\tiny{R}}}}{\dd W_t},\\
{\dd \overline{k}_t^{\text{\tiny{A}}}} &=-\sqrt{\lambda} \frac{a_t^{\text{\tiny{A}},\text{\tiny{I}}}}{a_t^{\text{\tiny{A}},\text{\tiny{R}}}}{\dd W_t}.
\end{split}
\end{equation}
Let us notice that the equation for $a_t^{\text{\tiny{A}}}$ is deterministic. The corresponding solution is reported in Ref.~\cite{Bassi2005}. The real ($a_t^{\text{\tiny{A}},\text{\tiny{R}}}$) and imaginary ($a_t^{\text{\tiny{A}},\text{\tiny{I}}}$) parts of $a_t^{\text{\tiny{A}}}$ read
\begin{equation} \label{atNL} 
\begin{split}
a_t^{\text{\tiny{A}},\text{\tiny{R}}}&=\frac{c^{\text{\tiny{R}}} \sinh[2(b^{\text{\tiny{R}}} t + k^{\text{\tiny{R}}})]-c^{\text{\tiny{I}}} \sin[2(b^{\text{\tiny{I}}} t + k^{\text{\tiny{I}}})]}{\cosh[2(b^{\text{\tiny{R}}} t+ k^{\text{\tiny{R}}})]+\cos[2(b^{\text{\tiny{I}}} t+ k^{\text{\tiny{I}}})]},\\
a_t^{\text{\tiny{A}},\text{\tiny{I}}}&=\frac{c^{\text{\tiny{I}}} \sinh[2 (b^{\text{\tiny{R}}} t + k^{\text{\tiny{R}}})+c^{\text{\tiny{R}}} \sin[2 (b^{\text{\tiny{I}}} t + k^{\text{\tiny{I}}})]]}{\cosh[2(b^{\text{\tiny{R}}} t + k^{\text{\tiny{R}}})]+\cos[2 (b^{\text{\tiny{I}}} t + k^{\text{\tiny{I}}})]},
\end{split}
\end{equation}
where $b=(1+i)\sqrt{\hbar \lambda/m}$, $c=(1-i)/2\sqrt{m\lambda/\hbar}$, $k~=~\tanh^{-1}[a_0/c]$ and $a_0$ is the value of $a_t^\text{\tiny{A}}$ at time $t=0$. 

Therefore, from Eq.~\eqref{Gaussianexps}, we see that we are now in position to determine $\Sigma^{\text{\tiny{A}}}_t(\hat{x})$. Thus, we have:
\begin{equation}\label{sigmaxA}
\Sigma^{\text{\tiny A}}_t(\hat{x})=\frac{1}{4}\frac{\cosh[2(b^{\text{\tiny{R}}} t+ k^{\text{\tiny{R}}})]+\cos[2(b^{\text{\tiny{I}}} t+ k^{\text{\tiny{I}}})]}{c^{\text{\tiny{R}}} \sinh[2(b^{\text{\tiny{R}}} t + k^{\text{\tiny{R}}})]-c^{\text{\tiny{I}}} \sin[2(b^{\text{\tiny{I}}} t + k^{\text{\tiny{I}}})]},
\end{equation}
Note that $\Sigma_t^{\text{\tiny{A}}}(\hat{x})$ does not depend on the noise, therefore the stochastic average is ineffective. We will see that the same happens for the linear unraveling.

Having solved for $a_t^{\text{\tiny{A}}}$, we can use the results of Eq.~\eqref{spreadcontmeas} to find the solution for $\overline{k}_t^{\text{\tiny{A}}}$. We obtain:
\begin{equation}
\overline{k}_t^{\text{\tiny{A}}}=\overline{k}_0 - \sqrt{\lambda}\int_0^t \frac{a_s^{\text{\tiny{A}},\text{\tiny{I}}}}{a_s^{\text{\tiny{A}},\text{\tiny{R}}}}\dd W_s. 
\end{equation}
Finally, from the dynamical evolution for  $\overline{x}_t^{\text{\tiny{A}}}$ in Eq.~\eqref{spreadcontmeas}, the previous results for $a_t^{\text{\tiny{A}}}$ and $\overline{k}_t^{\text{\tiny{A}}}$ lead to:
\begin{equation} \label{xtNL}
\begin{split}
\overline{x}_t^{\text{\tiny{A}}}&=\overline{x}_0 + \frac{\hbar}{m}\overline{k}_0 t - \sqrt{\lambda}\frac{\hbar}{m}\int_0^t (t-s)\frac{a_s^{\text{\tiny{A}},\text{\tiny{I}}}}{a_s^{\text{\tiny{A}},\text{\tiny{R}}}}\dd W_s\\
&+\frac{\sqrt{\lambda}}{2}\int_0^t \frac{1}{a_s^{\text{\tiny{A}},\text{\tiny{R}}}}\dd W_s.
\end{split}
\end{equation}

We are now in the position to solve Eq.~\eqref{x2NLdef}:
\begin{equation} \label{xt2NL}
\begin{split}
\mathbb{E}_\omega[({\ave{\hat{x}}_{t}^{\text{\tiny{A}}}})^2]
&=\left(\overline{x}_0+\frac{\hbar}{m}\overline{k}_0 t  \right)^2\\
&+ \lambda \int_0^t \dd s \frac{1}{(a_s^{\text{\tiny{A}},\text{\tiny{R}}})^2} \left(\frac{\hbar}{m}(t-s)a_s^{\text{\tiny{A}},\text{\tiny{I}}}-\frac{1}{2} \right)^2,
\end{split}
\end{equation}
which is the same result that one would obtain if by using the fact that $\ave{\hat{x}}_{t}^{\text{\tiny{A}}}=\overline{x}_t^{\text{\tiny{A}}}$ [cf. Eq.~\eqref{Gaussianexps}], and square and average the result of Eq.~\eqref{xtNL}.

\subsubsection{Linear unraveling}
\begin{figure}[t!]
    \centering
    \includegraphics[width=1\linewidth]{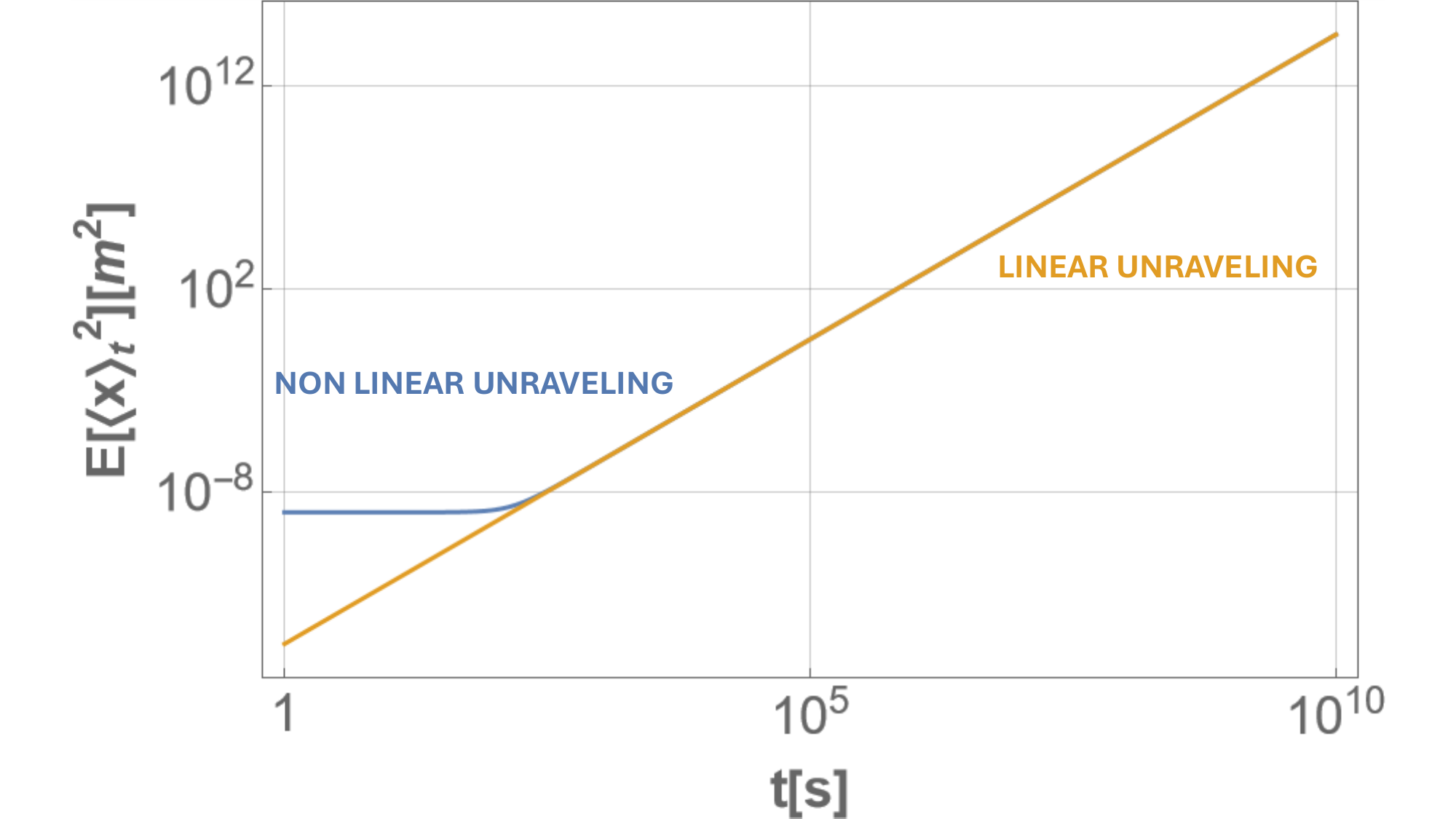}
    \caption{Stochastic average over all realizations of ${\ave{\hat{x}}_t}^2$, for the nonlinear (${\ave{\hat{x}}_t^{\text{\tiny{A}}}}^2$ in Eq.~\eqref{xt2NL}) and linear (${\ave{\hat{x}}_t^{\text{\tiny{B}}}}^2$ in Eq.~\eqref{xt2L}) unravelings. For illustrative purposes, we use the parameters: $\overline{x}_0~=~0$, $\overline{k}_0~=~0$,
    $m~=~10^{-15} \mathrm{kg}$, $a_0~=~0.25\times 10^{9} \mathrm{m^2}$, $\lambda~=~10^{23}\mathrm{m^{-2}\,Hz}$.}
    \label{LinearVSNONLinear}
\end{figure}

Let us now consider the linear unraveling of Eq.~\eqref{unitary}, with the same Gaussian ansatz of Eq.~\eqref{Gaussian}. In this case, the results of Eq.~\eqref{Gaussianexps} lead to the following evolution for ${\ave{\hat{x}}_{t}^{\text{\tiny{B}}}}^2$:
\begin{equation}
\dd {\ave{\hat{x}}_t^{\text{\tiny{B}}}}^2 = \frac{2 \hbar}{m}\overline{x}_t^{\text{\tiny{B}}} \overline{k}_t^{\text{\tiny{B}}} \dd t. 
\end{equation}
Working in the position representation, the evolution of the wave function corresponds to:
\begin{equation}
{\dd {\psi_t^{\text{\tiny{B}}}}(x)}=\left(\frac{i\hbar}{2m}\frac{\partial^2}{\partial x^2}{\dd t}-\frac{\lambda}{2}x^2 \dd t-i \sqrt{\lambda}{x}{\dd W_t}\right){\psi_t^{\text{\tiny{B}}}(x)},
\end{equation}
and the dynamical evolution for the relevant parameters of the Gaussian state reads:
\begin{equation} \label{paramunit}
\begin{split}
{\dd a_t^{\text{\tiny{B}}}}&= -\frac{2i\hbar}{m}{a_t^{\text{\tiny{B}}}}^2 {\dd t} ,\\
{\dd \overline{x}_t^{\text{\tiny{B}}}}&=\frac{\hbar}{m}\overline{k}_t^{\text{\tiny{B}}}{\dd t} ,\\
{\dd \overline{k}_t^{\text{\tiny{B}}}}&=-\sqrt{\lambda}{\dd W_t}.
\end{split}
\end{equation}
Solving for $\overline{k}_t^{\text{\tiny{B}}}$, we obtain: 
\begin{equation}
\overline{k}_t^{\text{\tiny{B}}}=\overline{k}_0 - \sqrt{\lambda} \int_0^t \dd W_s,    
\end{equation}
and the substitution of this result in the evolution of $\overline{x}_t^{\text{\tiny{B}}}$ in Eq.~\eqref{paramunit} yields to:
\begin{equation}
\overline{x}_t^{\text{\tiny{B}}}=\overline{x}_0 + \frac{\hbar}{m}\overline{k}_0 t-\frac{\hbar \sqrt{\lambda}}{m}\int_0^t (t-s) \dd W_s,
\end{equation}
with solution:
\begin{equation} \label{xt2L}
\begin{split}
\mathbb{E}_\omega[{\ave{\hat{x}}_{t}^{\text{\tiny{B}}}}^2 ]  =\left(\overline{x}_0 + \frac{\hbar}{m}\overline{k}_0 t  \right)^2 + \lambda \frac{\hbar^2}{m^2} \int_0^t (t-s)^2 \dd s,
\end{split}
\end{equation}
which in general differs from the expression for $\mathbb{E}_\omega[{\ave{\hat{x}}_{t}^{\text{\tiny{A}}}}^2]$, as shown in Fig. \ref{LinearVSNONLinear}.
\label{AppLinUnrav}
Finally, solving for $a_t^{\text{\tiny{B}}}$, we obtain 
\begin{equation} \label{spreadunit}
\begin{split}
a_t^{\text{\tiny{B}},\text{\tiny{R}}} &=\frac{m^2 a_0^R}{(m-2 \hbar t a_0^I)^2 + 4 \hbar^2 t^2 (a_0^R)^2},   \\
a_t^{\text{\tiny{B}},\text{\tiny{I}}} &=\frac{m^2 a_0^I-2 \hbar t|a_0|^2}{(m-2 \hbar t a_0^I)^2 + 4 \hbar^2 t^2 (a_0^R)^2}.
\end{split}
\end{equation}

Once again, from Eq.~\eqref{Gaussianexps}, we can calculate $\Sigma^{\text{\tiny{B}}}(\hat{x})$. We obtain:
\begin{equation} \label{sigmaxB}
\Sigma^{\text{\tiny B}}_t(\hat{x})=\frac{1}{4}\frac{(m-2 \hbar t a_0^{\text{\tiny{I}}})^2 + 4 \hbar^2 t^2 (a_0^{\text{\tiny{R}}})^2}{m^2 a_0^{\text{\tiny{R}}}},
\end{equation}
which is deterministic, as in the previous case.

\begin{figure}[t!]
    \centering
    \includegraphics[width=1\linewidth]{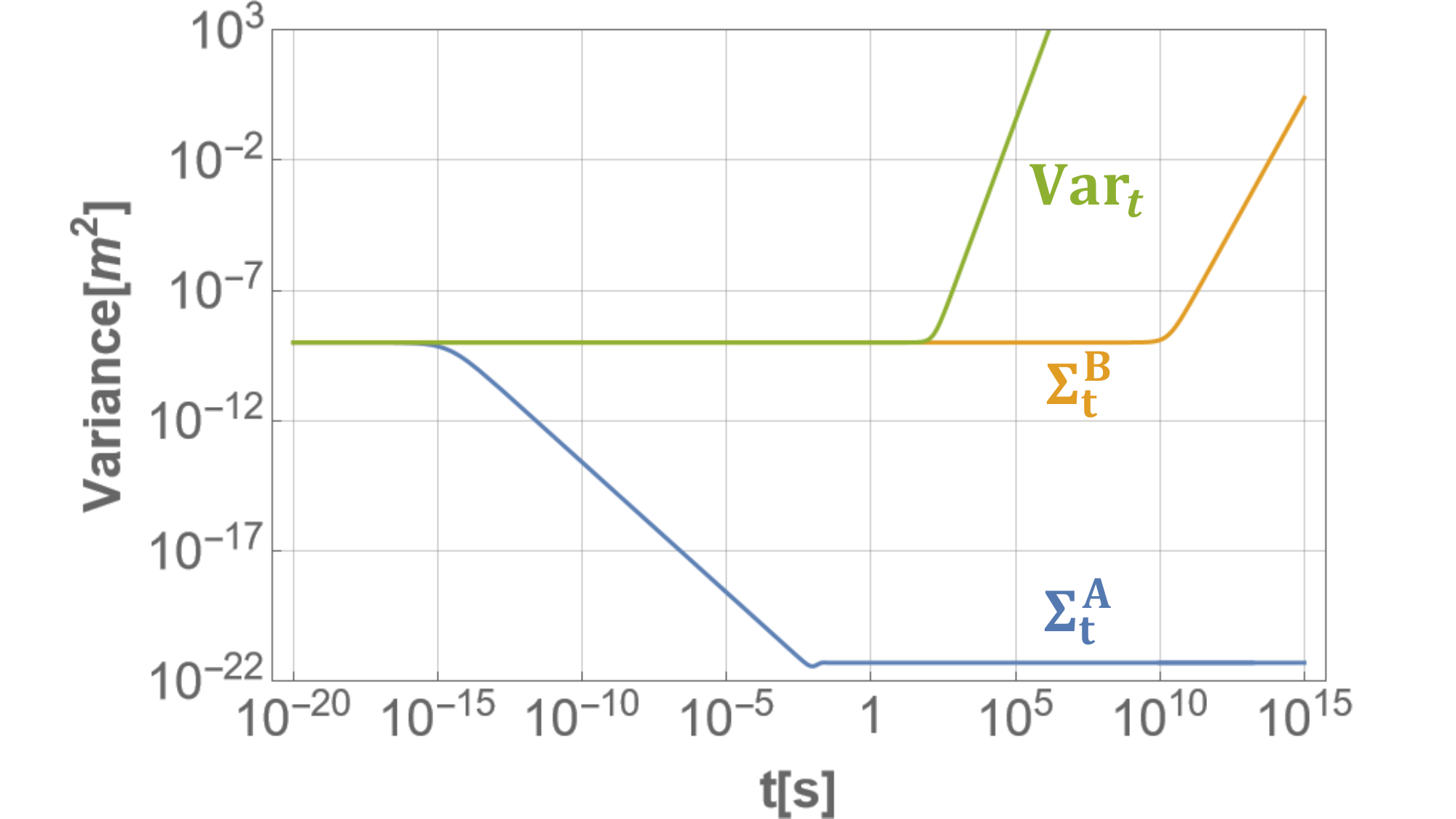}
    \caption{Time evolution of the unraveling-independent  variance $\text{Var}_t(\hat{x})$ (Eq.~\eqref{Varx}, green line), and of $\Sigma_t(\hat{x})$ for the nonlinear unraveling ($\Sigma_t^{\text{\tiny{A}}}(\hat{x})$ in Eq.~\eqref{sigmaxA}, blue line) and the linear unraveling ($\Sigma_t^{\text{\tiny{B}}}(\hat{x})$ in Eq.~\eqref{sigmaxB},  yellow line).
    For illustrative purposes, we use the parameters:
    $\overline{x}_0~=~0$, $\overline{k}_0~=~0$,
    $m~=~10^{-15} \, \mathrm{kg}$, $a_0~=~0.25\times 10^{9} \mathrm{m^{-2}}$ (therefore $\Sigma^{\text{\tiny A}}_0(\hat{x}) = \Sigma^{\text{\tiny B}}_0(\hat{x}) = \text{Var}_0(\hat{x}) = 10^{-9} \mathrm{m^{2}}$), $\lambda~=~10^{23}\mathrm{m^{-2}\,Hz}$. 
    The three quantities evolve differently over time.}
\label{LinearVSNONLinearVarianceReal}
\end{figure}

Both quantities $\Sigma_t^{\text{\tiny{A}}}(\hat{x})$ and $\Sigma_t^{\text{\tiny{B}}}(\hat{x})$ [cf. Eqs.~\eqref{sigmaxA} and ~\eqref{sigmaxB} ] differ from the  variance $\text{Var}_t(\hat{x})=\mathbb{E}_\omega[\ave{\hat{x}^2}_t]-(\mathbb{E}_\omega[\ave{\hat{x}}])^2_t =\text{Tr}[\hat{x}^2 \hat{\rho}_t]-(\text{Tr}[\hat{x}\hat{\rho}_t]
)^2$ \footnote{In general, we have $\mathbb{E}_\omega[\text{Var}(\hat{O}_i,\hat{O}_j)]=\frac{1}{2}\mathbb{E}_\omega\ave{\{\hat{O}_i,\hat{O}_j \}}_t-\mathbb{E}_\omega[\ave{\hat{O}_i}_t]\mathbb{E}_\omega[\ave{\hat{O}_j}_t]$.}, obtained from the density matrix, which by construction is unraveling-independent. Straightforward calculations show that 
\begin{equation} \label{Varx}
\begin{aligned}
    \text{Var}_t(\hat{x})&=\frac{(m-2 \hbar t a_0^{\text{\tiny{I}}})^2 + 4 \hbar^2 t^2 (a_0^{\text{\tiny{R}}})^2}{4m^2 a_0^{\text{\tiny{R}}}} + \lambda \frac{\hbar^2}{m^2} \frac{t^3}{3},
\end{aligned}
\end{equation}
which matches the the well-known expression  in the decoherence literature~\cite{joos2013decoherence} (for $a_0^{\text{\tiny{I}}}=0$). Fig.~\ref{LinearVSNONLinearVarianceReal} shows the time evolution of time of $\Sigma^{\text{\tiny A}}_t(\hat{x})$, $\Sigma^{\text{\tiny B}}_t(\hat{x})$, as well as of $\text{Var}_t(\hat{x})$, for given initial values of the parameters. For small times, they roughly coincide, because the free part of the dynamics, which is the same in all cases, dominates over the stochastic part. For larger times, the three quantities depart. $\Sigma^{\text{\tiny A}}_t(\hat{x})$ settles to a finite value: the nonlinear unraveling collapses the wave function, counteracting the free expansion, and therefore the Gaussian state reaches asymptotically a finite spread. The linear unraveling does not induce the collapse of the wave function, but acts as a random position-dependent potential, therefore the spread $\Sigma^{\text{\tiny B}}_t(\hat{x})$ keeps increasing over time. The same occurs for $\text{Var}_t(\hat{x})$ but at a faster rate. 
As a final remark, from the fact that the variance $\text{Var}_t(\hat{x})$ is unraveling-independent, it follows from Eq.~\eqref{Varx} and Eq.~\eqref{sigmaxA} that the term in the second line in Eq.~\eqref{xt2NL} is equal to:
\begin{equation}
    \begin{split}
        &\lambda \int_0^t \dd s \frac{1}{(a_s^{\text{\tiny{A}},\text{\tiny{R}}})^2} \left(\frac{\hbar}{m}(t-s)a_s^{\text{\tiny{A}},\text{\tiny{I}}}-\frac{1}{2} \right)^2=\\
        &=\frac{(m-2 \hbar t a_0^I)^2 + 4 \hbar^2 t^2 (a_0^R)^2}{4m^2 a_0^R} + \lambda \frac{\hbar^2}{m^2} \frac{t^3}{3}+\\
        &-\frac{1}{4}\frac{\cosh[2(b^R t + k^R)]+\cos[2 (b^I t + k^I)]}{c^I \sinh[2 (b^R t + k^R)+c^R \sin[2 (b^I t + k^I)]]}.
    \end{split}
\end{equation}

\subsection{Particle in a harmonic trap} 

Let us now consider a particle trapped in a harmonic potential:
\begin{equation}
\hat{H}=\frac{1}{2m}\hat{p}^2+\frac{1}{2}m \Omega^2 \hat{x}^2,
\end{equation}
and we maintain the choice $\hat{L}=\hat{x}$. We derive the dynamical evolution for both unravelings under the Gaussian ansatz of Eq.~\eqref{Gaussian}. 

\subsubsection{Nonlinear unraveling} 

\begin{figure}[h!] 
    \centering
    \includegraphics[width=1\linewidth]{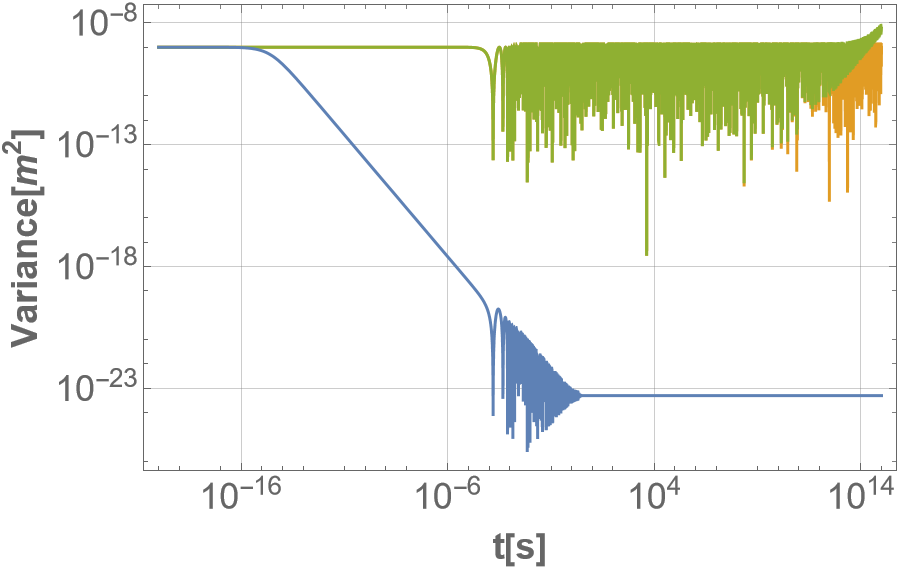} 
    \caption{Time evolution of the unraveling-independent variance $\text{Var}_t(\hat{x})$ (Eq.~\eqref{VarHarm}, green line), and of $\Sigma_t(\hat{x})$ for the nonlinear unraveling ($\Sigma_t^{\text{\tiny{A}}}(\hat{x})$ as in Eq.~\eqref{sigmaxA}, with $b$ and $c$ given in Eq.~\eqref{bHarm} and Eq.~\eqref{cHarm} [blue line]) and the linear unraveling ($\Sigma_t^{\text{\tiny{B}}}(\hat{x})$ as in Eq.~\eqref{sigmaxA}, with $b$ and $c$ given in Eq.~\eqref{bcharmlin} [yellow line]). For illustrative purposes, we use the parameters: $\overline{x}_0~=~0$, $\overline{k}_0~=~0$, $m~=~ 10^{-15} \, \mathrm{kg}$, $a_0~=~0.25\times 10^{9} \mathrm{m^{-2}}$, $\lambda~=~10^{23} \, \mathrm{m^{-2}\,Hz}$. The three quantities evolve differently over time. In addition here we assume $\Omega=10^4\,\mathrm{Hz}$ for the harmonic potential term.} 
\label{fig6}
\end{figure} 

We can follow the same procedure as in the previous section; the parameters of the Gaussian state $a_t^{\text{\tiny{A}}}$, $\overline{x}_t^{\text{\tiny{A}}}$ and $\overline{k}_t^{\text{\tiny{A}}}$ now obey the following dynamical evolution: 
\begin{equation} \label{atNLharm}
\begin{split}
    \dd a_t^{\text{\tiny{A}}}&=\left(\lambda + i \frac{m \Omega^2}{2\hbar} - \frac{2i \hbar}{m}{a_t^{\text{\tiny{A}}}}^2\right)\dd t,\\
    {\dd \overline{x}_t^{\text{\tiny{A}}}} &=\frac{\hbar}{m}\overline{k}_t^{\text{\tiny{A}}} \dd t+ \frac{\sqrt{\lambda} }{2 a_t^{\text{\tiny{A}},\text{\tiny{R}}}}{\dd W_t},\\
    {\dd \overline{k}_t^{\text{\tiny{A}}}}&=-\frac{m\Omega^2 }{\hbar}\overline{x}_t^{\text{\tiny{A}}}\dd t-\sqrt{\lambda} \frac{a_t^{\text{\tiny{A}},\text{\tiny{I}}}}{a_t^{\text{\tiny{A}},\text{\tiny{R}}}} {\dd W_t}.
\end{split} 
\end{equation}
The solutions for $\overline{x}_t^{\text{\tiny{A}}}$ and $\overline{k}_t^{\text{\tiny{A}}}$ read:
\begin{equation}
    \begin{split}
       &\overline{x}_t^{\text{\tiny{A}}}= \frac{\hbar
   }{m \Omega} \overline{k}_0\sin \left(\Omega t \right)+\overline{x}_0 \cos \left(\Omega t
   \right)+\\
   &+\sqrt{\lambda }\int_0^t \frac{2 a_t^{\text{\tiny{A}},\text{\tiny{I}}} \hbar \sin \left(\Omega (s-t)\right)+m \Omega \cos \left(\Omega(s-t)\right)}{2 a_t^{\text{\tiny{A}},\text{\tiny{R}}} m \Omega} \, \dd W_s,\\
   &\overline{k}_t^{\text{\tiny{A}}}=-\frac{m \Omega}{\hbar}\overline{x}_0 \sin(\Omega t)+\overline{k}_0
   \cos \left(\Omega t \right)+\\
   &+\sqrt{\lambda } \int_0^t \frac{m
   \Omega \sin \left(\Omega (s-t)\right)-2 a_t^{\text{\tiny{A}},\text{\tiny{I}}} \hbar \cos
   \left(\Omega (s-t)\right)}{2 a_t^{\text{\tiny{A}},\text{\tiny{R}}} \hbar} \, \dd W_s.
    \end{split}
\end{equation}
Thus, taking the square of $\overline{x}_t$ one gets:
\begin{equation} \label{xt2Aharm}
    \begin{split}
        &\mathbb{E}_\omega[{\ave{\hat{x}}_{t}^{\text{\tiny{A}}}}^2]=\left(\frac{\hbar}{m \Omega}\overline{k}_0 \sin(\Omega t)  +\overline{x}_0 \cos \left(\Omega t
   \right)\right)^2+\\
   &+\lambda \int_0^t \left(\frac{2 a_t^{\text{\tiny{A}},\text{\tiny{I}}} \hbar \sin
   \left(\Omega (s-t)\right)+m \Omega \cos \left(\Omega
   (s-t)\right)}{2 a_t^{\text{\tiny{A}},\text{\tiny{R}}} m \Omega} \right)^2\, ds.
    \end{split}
\end{equation}
The solution for $a_t^{\text{\tiny{A}}}$ is as in Eq.~\eqref{atNL}, where the parameters $b$ and $c$ are now defined as:
\begin{equation} \label{bHarm}
b=\frac{\hbar}{m}\sqrt{2(\sqrt{x_a^2+y_a^2}-x_a)}+i\frac{\hbar}{m}\sqrt{2\left(\sqrt{x_a^2+y_a^2}+x_a\right)},
\end{equation}
with $x_a=m^2 \Omega^2/(4\hbar^2)$ and $y_a=m \lambda/(2 \hbar)$. On the other hand, we have
\begin{equation} \label{cHarm}
c=\sqrt{\frac{1}{2}\left(\sqrt{x_a^2+y_a^2}+x_a \right)}-i \sqrt{\frac{1}{2}\left(\sqrt{x_a^2+y_a^2}-x_a \right)}. 
\end{equation}
We notice that in the limit in which $\Omega \rightarrow 0$, we have  that $x_a \rightarrow 0$ and thus we recover the definitions of the parameters $c$ and $b$ given after Eq.~\eqref{atNL}.

\subsubsection{Linear unraveling}

The stochastic differential equations for the Gaussian state parameters $a_t^{\text{\tiny{B}}}$, $\overline{x}_t^{\text{\tiny{B}}}$ and $\overline{k}_t^{\text{\tiny{B}}}$ are given by:
\begin{equation}
\begin{split}
{\dd a_t^{\text{\tiny{B}}}}&=\left(i \frac{m \Omega^2}{2 \hbar}-\frac{2i \hbar}{m}{a_t^{\text{\tiny{B}}}}^2\right)\dd t,\\
{\dd \overline{x}_t^{\text{\tiny{B}}}}&=\frac{\hbar}{m}\overline{k}_t^{\text{\tiny{B}}}{\dd t},\\
{\dd \overline{k}_t^{\text{\tiny{B}}}}&=-\frac{m \Omega^2}{\hbar}\overline{x}_t^{\text{\tiny{B}}}{\dd t} - \sqrt{\lambda} {\dd W_t}.
\end{split}
\end{equation}
The above equations lead to the following solutions for  $\overline{x}_t^{\text{\tiny{B}}}$ and $\overline{k}_t^{\text{\tiny{B}}}$:
\begin{equation}
    \begin{split}
    &\overline{x}_t^{\text{\tiny{B}}}\!\!=\!\!\frac{\hbar}{m \Omega}\overline{k}_0 \sin(\Omega t)\! +\!\overline{x}_0 \cos \left(\Omega t
   \right)\!+\!\frac{\sqrt{\lambda}\hbar}{m \Omega}\!\! \int_0^t \! \! \sin(\Omega(s-t))  \dd W_s,\\
   &\overline{k}_t^{\text{\tiny{B}}}\! \!=\! \!-\frac{m \Omega}{\hbar}\overline{x}_0 \sin(\Omega t)\!+\!\overline{k}_0
   \cos \left(\Omega t \right)\!-\!\sqrt{\lambda } \!\int_0^t \! \!  \cos
   \left(\Omega (s-t)\right)\! \dd W_s,
    \end{split}
\end{equation}
and it follows that:
\begin{equation} \label{xt2Bharm}
    \begin{split}
  &\mathbb{E}_\omega[{\ave{\hat{x}}_{t}^{\text{\tiny{B}}}}^2]=\left(\frac{\hbar
  }{m \Omega } \overline{k}_0 \sin \left(\Omega t \right)+\overline{x}_0 \cos \left(\Omega t
   \right)\right)^2+\\
   &+\int_0^t \left(\frac{\sqrt{\lambda }  \hbar \sin
   \left(\Omega (s-t)\right)}{  m \Omega}\right)^2\, \dd s.
    \end{split}
\end{equation}
We notice that the equation for $a_t^{\text{\tiny{B}}}$ is the particular case $\lambda=0$ of Eq.~\eqref{atNLharm}. Therefore, the solution of this equation can be readily obtained from Eq.~\eqref{atNL}, where we set $y_a=0$ and find
\begin{equation} \label{bcharmlin}
b=i \Omega, \qquad c=\frac{m \Omega}{2 \hbar}.
\end{equation}
Finally, the unraveling-independent variance $\text{Var}_t(\hat{x})$ is given by
\begin{equation} \label{VarHarm}
\begin{split}
\text{Var}_t(\hat{x})&=\frac{\hbar}{2m \Omega}\frac{\cosh[2k^\text{\tiny{R}}]+\cos[2 (\Omega t + k^{\text{\tiny{I}}})]}{\sinh[2 k^{\text{\tiny{R}}}]} \\
&+ \frac{\lambda \hbar^2}{2m^2 \Omega^2}\left(t-\frac{\sin(2 \Omega t)}{2 \Omega} \right), 
\end{split}
\end{equation}
with
\begin{equation}
k^{\text{\tiny{R}}}=\frac{1}{4}\ln(x_v^2+y_v^2), \qquad k^{\text{\tiny{I}}}=\frac{1}{2}\arctan \left(\frac{y_v}{x_v} \right),    
\end{equation}
where
\begin{equation}
\begin{split}
x_v&=\frac{|c|^2-|a_0|^2}{(c^{\text{\tiny{R}}}-a_0^{\text{\tiny{R}}})^2+(c^{\text{\tiny{I}}}-a_0^{\text{\tiny{I}}})^2}, \\
y_v&=\frac{2 (a_0^{\text{\tiny{I}}}c^{\text{\tiny{R}}}-a_0^{\text{\tiny{R}}}c^{\text{\tiny{I}}})}{(c^{\text{\tiny{R}}}-a_0^{\text{\tiny{R}}})^2+(c^{\text{\tiny{I}}}-a_0^{\text{\tiny{I}}})^2}.    
\end{split}
\end{equation}
and the parameters $a_0$ and $c$ are those corresponding to the linear unraveling. In analogy with the free particle case, Fig.~\ref{fig6} shows the evolution of $\Sigma_t^{\tiny{A}}(\hat{x})$, $\Sigma_t^{\tiny{B}}(\hat{x})$ and $\text{Var}_t(\hat{x})$ for the particle in the harmonic trap.

\vspace{2 cm}
\bibliography{biblio}

@article{Bassi2005,
  title={Collapse models: analysis of the free particle dynamics},
  author={A. Bassi},
  journal={J. Phys. A: Math. Gen.},
  volume={38},
  pages={3173},
  year={2005},
  DOI={10.1088/0305-4470/38/14/008}
}

@article{Vovk2022,
  title={Entanglement-Optimal Trajectories of Many-Body Quantum Markov Processes},
  author={T. Vovk and H. Pichler},
  journal={Phys. Rev. Lett.},
  volume={128},
  pages={243601},
  year={2022},
  DOI={https://doi.org/10.1103/PhysRevLett.128.243601}
}

@article{Gorini1976,
  title={Completely positive dynamical semigroups of N‐level systems},  
  author={V. Gorini and A. Kossakowski and E. C. G. Sudarshan},
  journal={J. Math. Phys.},
  volume={17},
  pages={821},
  year={1976},
  DOI={https://doi.org/10.1063/1.522979},
}

@article{Lindblad1976,
  title={On the generators of quantum dynamical semigroups},
  author={G. Lindblad},
  journal={Commun. Math. Phys.},
  volume={48},
  pages={119},
  year={1976},
  DOI={https://doi.org/10.1007/BF01608499},
}

@article{Caiaffa2017,
  title={Stochastic unraveling of positive quantum dynamics},
  author={M. Caiaffa and A. Smirne and A. Bassi},
  journal={Phys. Rev. A},
  volume={95},
  pages={062101},
  year={2017},
  DOI={https://doi.org/10.1103/PhysRevA.95.062101},
}

@article{Donvil2023,
  title={On the Unraveling of Open Quantum Dynamics},
  author={Donvil, B. I. C. and Muratore-Ginanneschi, P.},
  journal={Open Syst. Inf. Dyn.},
  volume={30},
  number={},
  pages={2350015},
  year={2023},
  publisher={World Scientific},
  DOI={https://doi.org/10.1142/S1230161223500154}
}

@article{Gardiner1992,
  title={Wave-function quantum stochastic differential equations and quantum-jump simulation methods},
  author={Gardiner, C.W. and Parkins, A. S. and Zoller, P.},
  journal={Phys. Rev. A},
  volume={46},
  number={},
  pages={4363},
  year={1992},
  publisher={APS},
  DOI={https://doi.org/10.1103/PhysRevA.46.4363},
}

@article{Kondov2003,
  title={Stochastic unraveling of Redfield master equations and its application to electron transfer problems},
  author={Kondov, I. and Kleinekath\"ofer, U. and Schreiber, M.},
  journal={J. Chem. Phys.},
  volume={119},
  number={},
  pages={6635},
  year={2003},
  publisher={AIP Publishing},
  DOI={https://doi.org/10.1063/1.1605095}
}

@article{Li2013,
  title={Simulation of Quantum Dynamics Based on the Quantum Stochastic Differential Equation},
  author={Li, M.},
  journal={Sci. World J.},
  volume={2013},
  number={},
  pages={424137},
  year={2013},
  publisher={Wiley},
  DOI={https://doi.org/10.1155/2013/424137}
}

@article{Piñol2024,
  title={Telling different unravelings apart via nonlinear quantum-trajectory averages},
  author={E. Piñol and Th. K. Mavrogordatos and D. Keys and R. Veyron and P. Sierant and M. A. Garc\'ia-March and S. Grandi and M. W. Mitchell and J. Wehr and M. Lewenstein },
  journal={Phys. Rev. Research},
  volume={6},
  pages={L032057},
  year={2024},
  DOI={https://doi.org/10.1103/PhysRevResearch.6.L032057},
}

@article{Piccitto2024,
  title={The impact of different unravelings in a monitored system of free fermions},
  author={G. Piccitto and D. Rossini and A. Russomanno},
  journal={Eur. Phys. J. B},
  volume={97},
  pages={90},
  year={2024},
  DOI={https://doi.org/10.1140/epjb/s10051-024-00725-0}
}

@article{Wu2024,
  title={Squeezing below the ground state of motion of a continuously monitored levitating nanoparticle},
  author={Q. Wu and D. A. Chisholm and R. Muffato and T. Georgescu and J. Homans and H. Ulbricht and M. Carlesso and M. Paternostro},
  journal={Quantum Sci. Technol.},
  volume={9},
  pages={045038},
  year={2024},
  DOI={10.1088/2058-9565/ad7284},
}

@book{Jacobs2010,
    author={K. Jacobs},
    title ={Stochastic Processes for Physicists. Understanding Noisy Systems} ,
    publisher = {Cambridge University Press} ,
    year = {2010}, 
}

@book{jacobs2014quantum,
  title={Quantum measurement theory and its applications},
  author={Jacobs, Kurt},
  year={2014},
  publisher={Cambridge University Press}
}

@article{Genoni2016,
  title={Conditional and unconditional Gaussian quantum dynamics},
  author={M. G. Genoni and L. Lami and A. Serafini},
  journal={Contemp. Phys.},
  volume={57},
  pages={331},
  year={2016},
  DOI={https://doi.org/10.1080/00107514.2015.1125624},
}

@article{Cao2019,
  title={Entanglement in a fermion chain under continuous monitoring},
  author={X. Cao and A. Tilloy and A. De Luca},
  journal={SciPost Phys.},
  volume={7},
  pages={024},
  year={2019},
  DOI={10.21468/SciPostPhys.7.2.024}
}

@book{breuer2002theory,
  title={The theory of open quantum systems},
  author={Breuer, Heinz-Peter and Petruccione, Francesco},
  year={2002},
  publisher={Oxford University Press, USA}
}

@book{carmichael2009open,
  title={An open systems approach to quantum optics: Lectures presented at the Universit{\'e} Libre de Bruxelles, October 28 to November 4, 1991},
  author={Carmichael, Howard},
  volume={18},
  year={2009},
  publisher={Springer Science \& Business Media}
}

@article{daley2014quantum,
  title={Quantum trajectories and open many-body quantum systems},
  author={Daley, Andrew J},
  journal={Adv. Phys.},
  volume={63},
  number={2},
  pages={77--149},
  year={2014},
  publisher={Taylor \& Francis},
  DOI={https://doi.org/10.1080/00018732.2014.933502},
}

@article{dalibard1992wave,
  title={Wave-function approach to dissipative processes in quantum optics},
  author={Dalibard, Jean and Castin, Yvan and M{\o}lmer, Klaus},
  journal={Phys. Rev. Lett.},
  volume={68},
  number={5},
  pages={580},
  year={1992},
  publisher={APS},
  DOI={https://doi.org/10.1103/PhysRevLett.68.580}
}

@article{gisin1992quantum,
  title={The quantum-state diffusion model applied to open systems},
  author={Gisin, Nicolas and Percival, Ian C},
  journal={J. Phys. A: Math. Gen},
  volume={25},
  number={21},
  pages={5677},
  year={1992},
  publisher={IOP Publishing},
  DOI={10.1088/0305-4470/25/21/023},
}

@article{hanggi1990reaction,
  title={Reaction-rate theory: fifty years after Kramers},
  author={H{\"a}nggi, Peter and Talkner, Peter and Borkovec, Michal},
  journal={Rev. Mod. Phys.},
  volume={62},
  number={2},
  pages={251},
  year={1990},
  publisher={APS},
  DOI={https://doi.org/10.1103/RevModPhys.62.251},
}

@book{nielsen2010quantum,
  title={Quantum computation and quantum information},
  author={Nielsen, Michael A and Chuang, Isaac L},
  year={2010},
  publisher={Cambridge University Press}
}

@article{molmer1993monte,
  title={Monte Carlo wave-function method in quantum optics},
  author={M{\o}lmer, Klaus and Castin, Yvan and Dalibard, Jean},
  journal={JOSA B},
  volume={10},
  number={3},
  pages={524--538},
  year={1993},
  publisher={Optica Publishing Group},
  DOI={https://doi.org/10.1364/JOSAB.10.000524}
}

@article{preskill2018quantum,
  title={Quantum computing in the NISQ era and beyond},
  author={Preskill, John},
  journal={Quantum},
  volume={2},
  pages={79},
  year={2018},
  publisher={Verein zur F{\"o}rderung des Open Access Publizierens in den Quantenwissenschaften},
  DOI={https://doi.org/10.22331/q-2018-08-06-79},
}

@book{wiseman2009quantum,
  title={Quantum measurement and control},
  author={Wiseman, Howard M and Milburn, Gerard J},
  year={2009},
  publisher={Cambridge University Press}
}

@book{weiss2012quantum,
  title={Quantum dissipative systems},
  author={Weiss, Ulrich},
  year={2012},
  publisher={World Scientific}
}

@article{wiseman1993quantum,
  title={Quantum theory of field-quadrature measurements},
  author={Wiseman, Howard M and Milburn, Gerard J},
  journal={Phys. Rev. A},
  volume={47},
  number={1},
  pages={642},
  year={1993},
  publisher={APS},
  DOI={https://doi.org/10.1103/PhysRevA.47.642},
}

@article{piccitto2022entanglement,
  title={Entanglement transitions in the quantum Ising chain: A comparison between different unravelings of the same Lindbladian},
  author={Piccitto, Giulia and Russomanno, Angelo and Rossini, Davide},
  journal={Phys. Rev. B},
  volume={105},
  number={6},
  pages={064305},
  year={2022},
  publisher={APS},
  DOI={https://doi.org/10.1103/PhysRevB.105.064305}
}

@article{vovk2024quantum,
  title={Quantum trajectory entanglement in various unravelings of Markovian dynamics},
  author={Vovk, Tatiana and Pichler, Hannes},
  journal={Phys. Rev. A},
  volume={110},
  number={1},
  pages={012207},
  year={2024},
  publisher={APS},
  DOI={https://doi.org/10.1103/PhysRevA.110.012207}
}

@book{joos2013decoherence,
  title={Decoherence and the appearance of a classical world in quantum theory},
  author={Joos, Erich and Zeh, H Dieter and Kiefer, Claus and Giulini, Domenico JW and Kupsch, Joachim and Stamatescu, Ion-Olimpiu},
  year={2013},
  publisher={Springer Science \& Business Media}
}

@article{wiseman1996quantum,
  title={Quantum trajectories and quantum measurement theory},
  author={Wiseman, Howard M},
  journal={Quantum Semiclass. Opt.},
  volume={8},
  number={1},
  pages={205},
  year={1996},
  publisher={IOP Publishing},
  DOI={10.1088/1355-5111/8/1/015},
}

@article{bassi2006geometric,
  title={Geometric phase for open quantum systems and stochastic unravelings},
  author={Bassi, Angelo and Ippoliti, Emiliano},
  journal={Phys. Rev. A},
  volume={73},
  number={6},
  pages={062104},
  year={2006},
  publisher={APS},
  DOI={https://doi.org/10.1103/PhysRevA.73.062104},
}

@inproceedings{carmichael2007quantum,
  title={Quantum jumps revisited: An overview of quantum trajectory theory},
  author={Carmichael, HJ},
  booktitle={Quantum Future From Volta and Como to the Present and Beyond: Proceedings of the Xth Max Born Symposium Held in Przesieka, Poland, 24--27 September 1997},
  pages={15--36},
  year={2007},
  organization={Springer}
}

@article{magrini2021real,
  title={Real-time optimal quantum control of mechanical motion at room temperature},
  author={Magrini, Lorenzo and Rosenzweig, Philipp and Bach, Constanze and Deutschmann-Olek, Andreas and Hofer, Sebastian G and Hong, Sungkun and Kiesel, Nikolai and Kugi, Andreas and Aspelmeyer, Markus},
  journal={Nature},
  volume={595},
  number={7867},
  pages={373--377},
  year={2021},
  publisher={Nature Publishing Group UK London},
  DOI={https://doi.org/10.1038/s41586-021-03602-3},
}

@article{albarelli2017ultimate,
  title={Ultimate limits for quantum magnetometry via time-continuous measurements},
  author={Albarelli, Francesco and Rossi, Matteo AC and Paris, Matteo GA and Genoni, Marco G},
  journal={New J. Phys.},
  volume={19},
  number={12},
  pages={123011},
  year={2017},
  publisher={IOP Publishing},
  DOI={10.1088/1367-2630/aa9840},
}

@article{Diosi1988,
  title={A universal master equation for the gravitational violation of quantum mechanics},
  author={Di\'osi, Lajos},
  journal={Phys. Lett. A},
  volume={120},
  number={8},
  pages={377-381},
  year={1998},
  publisher={Elsevier},
  DOI={https://doi.org/10.1016/0375-9601(87)90681-5},
}

@article{Ghirardi1990,
  title={Markov processes in Hilbert space and continuous spontaneous localization of systems of identical particles},
  author={Ghirardi, GianCarlo and Pearle, Philip and Rimini, Alberto},
  journal={Phys. Rev. A},
  volume={42},
  number={},
  pages={78},
  year={1990},
  publisher={APS},
  DOI={https://doi.org/10.1103/PhysRevA.42.78}
}

@article{Diosi2015,
  title={Is spontaneous wave function collapse testable at all?},
  author={Diósi, L.},
  journal={J. Phys.: Conf. Ser.},
  volume={626},
  pages={012008},
  year={2015},
  publisher={IOP},
  DOI={10.1088/1742-6596/626/1/012008},
}

@inbook{Diosi2018,
    author ={Diósi, L.} ,
    title ={How to Teach and Think About Spontaneous Wave Function Collapse Theories: Not Like Before} ,
    booktitle={Collapse of the Wave Function.
Models, Ontology, Origin, and Implications},
    publisher ={Cambridge University Press} ,
    year = {2018},
    chapter = {1},
}

@article{Kafri2014,
  title={A classical channel model for gravitational decoherence},
  author={Kafri, D. and Taylor, J. M. and Milburn, G. J.},
  journal={New J. Phys.},
  volume={16},
  number={},
  pages={065020},
  year={2014},
  publisher={IOP},
  DOI={10.1088/1367-2630/16/6/065020},
}

@article{Tilloy2016,
  title={Sourcing semiclassical gravity from spontaneously localized quantum matter},
  author={Tilloy, Antoine and Di\'osi, Lajos},
  journal={Phys. Rev. D},
  volume={93},
  number={},
  pages={024026},
  year={2016},
  publisher={APS},
  DOI={https://doi.org/10.1103/PhysRevD.93.024026},
}

@article{Tilloy2017,
  title={Principle of least decoherence for Newtonian semiclassical gravity},
  author={Tilloy, Antoine and Di\'osi, Lajos},
  journal={Phys. Rev. D},
  volume={96},
  number={},
  pages={104045},
  year={2017},
  publisher={APS},
  DOI={https://doi.org/10.1103/PhysRevD.96.104045},
}

@article{Wiseman2001,
  title={Complete parameterization, and invariance, of diffusive quantum trajectories for Markovian open systems},
  author={Wiseman, H. M. and Di\'osi, L.},
  journal={Chem. Phys.},
  volume={268},
  number={},
  pages={91},
  year={2001},
  publisher={Elsevier},
  DOI={https://doi.org/10.1016/S0301-0104(01)00296-8},
}

@article{zhang2017prediction,
  title={Prediction and retrodiction with continuously monitored Gaussian states},
  author={Zhang, Jinglei and M{\o}lmer, Klaus},
  journal={Phys. Rev. A},
  volume={96},
  number={6},
  pages={062131},
  year={2017},
  publisher={APS},
  DOI={https://doi.org/10.1103/PhysRevA.96.062131},
}

@article{belenchia2020entropy,
  title={Entropy production in continuously measured Gaussian quantum systems},
  author={Belenchia, Alessio and Mancino, Luca and Landi, Gabriel T and Paternostro, Mauro},
  journal={npj Quantum Inf.},
  volume={6},
  number={1},
  pages={97},
  year={2020},
  publisher={Nature Publishing Group UK London},
  DOI={https://doi.org/10.1038/s41534-020-00334-6},
}

@article{zhang2017quantum,
  title={Quantum feedback: theory, experiments, and applications},
  author={Zhang, Jing and Liu, Yu-xi and Wu, Re-Bing and Jacobs, Kurt and Nori, Franco},
  journal={Phys. Rep.},
  volume={679},
  pages={1--60},
  year={2017},
  publisher={Elsevier},
  DOI={https://doi.org/10.1016/j.physrep.2017.02.003},
}

@article{Adler2007,
  title={Collapse models with non-white noises},
  author={Adler, Stephen L. and Bassi, Angelo},
  journal={J. Phys. A: Math. Theor},
  volume={40},
  number={50},
  pages={15083},
  year={2007},
  publisher={IOP},
  DOI={10.1088/1751-8113/40/50/012}
}

@article{Bassi2003,
  title={Dynamical reduction models},
  author={Bassi, Angelo and Ghirardi, GianCarlo},
  journal={Phys. Rep.},
  volume={379},
  number={5-6},
  pages={257},
  year={2003},
  publisher={Elsevier},
  DOI={https://doi.org/10.1016/S0370-1573(03)00103-0},
}

@article{Bassi2013,
  title={Models of wave-function collapse, underlying theories, and experimental tests},
  author={Bassi, Angelo and Lochan, Kinjalk and Satin, Seema and Singh, Teijinder P. and Ulbricht, Hendrik},
  journal={Rev. Mod. Phys.},
  volume={85},
  number={},
  pages={471},
  year={2013},
  publisher={APS},
  DOI={https://doi.org/10.1103/RevModPhys.85.471},
}

@book{Rivas2012,
  title={Open Quantum Systems. An Introduction},
  author={Rivas, Angel and Huelga, Susana F.},
  year={2012},
  publisher={Springer Berlin}
}

@article{Ghirardi1990b,
  title={Relativistic dynamical reduction models: General framework and examples},
  author={Ghirardi, G.C. and Grassi, R. and Pearle, P.},
  journal={Found. Phys.},
  volume={20},
  number={},
  pages={1271},
  year={1990},
  publisher={Springer},
  DOI={https://doi.org/10.1007/BF01883487},
}

@inproceedings{Pearle1997,
    author = {Pearle, Philip} ,
    title ={True Collapse and False Collapse},
    booktitle = {Quantum classical correspondence : Proceedings of the 4th Drexel Symposium on Quantum Nonintegrability, Drexel University, Philadelphia, USA, September 8-11, 1994 / edited by Da Hsuan Feng, Bei Lok Hu.} ,
pages={51},
year = {1997},
publisher={Cambridge, MA: International Press}
}

@article{Albarelli2024,
  title={A pedagogical introduction to continuously monitored quantum systems and measurement-based feedback},
  author={Albarelli, F. and Genoni, M. G. },
  journal={Phys. Lett. A},
  volume={494},
  number={},
  pages={129260},
  year={2024},
  publisher={Elsevier},
  DOI={https://doi.org/10.1016/j.physleta.2023.129260},
}

@phdthesis{Wiseman1994,
    author={Wiseman, H. M.} ,
    title={Quantum Trajectories and Feedback},
    school ={University of Queensland} ,
    year ={1994} 
}

@book{Gardiner2000,
  title={Quantum Noise. A Handbook of Markovian and Non-Markovian Quantum Stochastic Methods with Applications to Quantum Optics},
  author={Gardiner, C. W. and Zoller, P.},
  year={2000},
  publisher={Springer-Verlag Berlin Heidelberg}
}

@article{Setter2018,
  title={Real-time Kalman filter: Cooling of an optically levitated nanoparticle},
  author={Setter, A. and Toroš, M. and Ralph, J. F. and Ulbricht, H.},
  journal={Phys. Rev. A},
  volume={97},
  pages={033822},
  year={2018},
  DOI={https://doi.org/10.1103/PhysRevA.97.033822}
}

\end{document}